\documentclass[12pt,a4paper]{article}
\pdfoutput=1
 \usepackage{jheppub}
 \usepackage{amsmath}
 \usepackage{amsfonts}
 \usepackage{mathtools}
 \usepackage{amssymb}
 \usepackage{caption}
 \usepackage{subcaption}
 \usepackage{slashed}
\usepackage{float}
\usepackage{graphicx}
\graphicspath{ {project2 tex/} }

\def\n{\nu}

\def\k{\kappa}

\newcommand{\be} {\begin{equation}}
\newcommand{\ee} {\end{equation}}
\newcommand{\bea} {\begin{eqnarray}}
\newcommand{\eea} {\end{eqnarray}}
\newcommand{\ba} {\begin{array}}
\newcommand{\ea} {\end{array}}
\newcommand{\nn} {\nonumber}

 \title{ Properties of dyons  in  ${\cal N}=4$ theories  at small charges}
 \author{Aradhita Chattopadhyaya, Justin R. David}
\affiliation{Centre for High Energy Physics, Indian Institute of Science,\\
C. V. Raman Avenue, Bangalore 560012, India.}
\emailAdd{aradhitac, justin@iisc.ac.in}

\abstract{
We study  three properties of $1/4$ BPS dyons at small charges in string compactifications which preserve
${\cal N}=4$ supersymmetry.  
We evaluate the non-trivial constant present in the one loop statistical entropy 
 for   ${\cal N}=4$ compactifications  of type IIB theory on 
$K3\times T^2$ orbifolded by  an order $\mathbb{Z}_N$ freely acting orbifold   $g'$ including all CHL compactifications. 
 This constant is trivial  for the un-orbifolded model but  we show that  it contributes crucially  to the entropy of low charge 
 dyons in all the orbifold  models.
 We then show that the  meromorphic Jacobi form which captures the degeneracy of $1/4$ BPS states for 
  the first two non-trivial magnetic charges 
  can be decomposed into an Appell-Lerch sum and 
 a mock  Jacobi form transforming under $\Gamma_0(N)$.  
 This generalizes the earlier observation of Dabholkar-Murthy-Zagier to the orbifold models.  
 Finally we study the sign of the Fourier coefficients of the inverse Siegel modular form which counts 
 the index of  $1/4$ BPS dyons  in ${\cal N}=4$ models obtained by  freely acting $\mathbb{Z}_2$ and 
 $\mathbb{Z}_3$ orbifolds of type II theory compactified on $T^6$. 
 We show that sign  of the index for sufficiently low charges  and ensuring that it counts single centered 
 black holes,  violates the positivity conjecture of Sen which indicates that these states posses non-trivial hair. 
}

\begin{document}
\maketitle
\flushbottom

\section{Introduction}

One of the successes of string theory as a  theory of quantum gravity lies in its 
microscopic  understanding of 
the Bekenstein-Hawking entropy. In \cite{Strominger:1996sh} it was shown that the 
statistical entropy of  a system of branes  carrying the same quantum numbers of 
the black hole agrees precisely with that of a class of extremal black holes in 5 dimensions. 
In 4 dimensions, 
starting with the  original work of \cite{Dijkgraaf:1996it}, the degeneracies of $1/4$ BPS dyons  in 
heterotic string theory compactified on $T^6$  and its
generalizations to  CHL compactifications \cite{Jatkar:2005bh} provide examples where a precise
formula for the microscopic degeneracies of extremal black holes in known. 
The degeneracy of dyons or more precisely  an index  can be obtained 
from the Fourier coefficients of  the inverse of an appropriate $Sp(2, \mathbb{Z})$ 
Siegel modular form. 
On taking the large charge limit of the index, the logarithm of the index agrees precisely 
not only with the Bekenstein-Hawking entropy of the corresponding black hole with the 
same charges, but also with the sub-leading correction given by the Wald's  generalization of 
the Bekenstein-Hawking formula \cite{LopesCardoso:2004law,David:2006yn}
\footnote{See \cite{Sen:2007qy, Dabholkar:2012zz} for reviews.}.

Given the success of the microscopic formula for the degeneracies of dyons in the large 
charge limit, it is natural to study its properties for dyons with small charges. 
This will hopefully provide more intuition on how a geometric description for the dyons 
in terms of metric can arise. 
In this paper we study three properties  $1/4$ BPS dyons in a class of 
${\cal N} = 4$ compactifications of string theory.  
These compactifications arise from considering type IIB theory on $K3\times T^2$ orbifolded by 
by $g'$ which acts as an $\mathbb{Z}_N$ automorphism on $K3$ together with a 
$1/N$ shift on one of the circles of $T^2$. 
This class includes all the CHL compactifications as well as more general ones where 
$g'$ corresponds to conjugacy classes of the Mathieu group $M_{23}$.  
The orbifolds we  study are listed table \ref{tt}.  The partition function of dyons in terms 
of inverse of a Siegel modular forms for these compactifications were constructed in 
\cite{Persson:2013xpa,Chattopadhyaya:2017ews}. 
We also study certain  $\mathbb{Z}_2, \mathbb{Z}_3$ compactifications of type II theory on 
$T^6$. 

 The degeneracy in the large charge limit 
is obtained by evaluating the leading  saddle point of an integral  and its one loop correction. 
The integral   extracts out the 
Fourier coefficient of the  inverse  Siegel modular form.  Let us call the logarithm of this saddle point 
approximation for the degeneracy  $S^{(1)}_{\rm stat} (Q, P)$ for 
definiteness where $Q$ and $P$ are the electric and magnetic charges of the dyon. 
 It is $S^{(1)}_{\rm stat}$ that agrees with the Bekenstein-Hawking formula  and its
generalization by Wald evaluated on a dyonic extremal black hole with the same set of charges. 
In \cite{Sen:2007qy} and \cite{Banerjee:2008ky}, $S^{(1)}_{\rm sat}$ 
was compared with the exact degeneracy 
of dyons 
at  low charges \footnote{These charges were such that the dyons are single centered.}.
 For the ${\cal N}=4$   heterotic string theory on $T^6$ or equivalently 
type IIB theory on $K3\times T^2$
it was seen that $S^{(1)}_{\rm sat}$  remarkably agrees with the exact degeneracy to within 
$2\%$ even for the lowest admissible charge. 
We re-evaluate $S^{(1)}_{\rm sat}$ in this paper for all ${\cal N} =4$ compactifications
arising as $g'$  orbifolds of $K3\times T^2$ with $g'$ given in table \ref{tt} keeping 
track of a constant $C_1$  in $S^{(1)}_{\rm sat}$.  This constant is trivial that is,  
$\ln C_1 =0$,  for the un-orbifolded compactification on $K3\times T^2$ and was not determined earlier 
for the orbifolds.
We then compare the statistical entropy at one loop $S^{(1)}_{\rm sat}$ to the exact degeneracy 
and show that the constant plays a crucial rule. In fact  without  $C_1$,   the agreement
with the exact degeneracy at low charges is off by more that $50\%$ in most cases. 
The reader can directly go to tables \ref{tt2a} to \ref{tt15a}  to appreciate the presence of $C_1$  in $S^{(1)}_{\rm sat}$.
The constant $C_1$ also  contributes to all the subsequent saddle points. 
To demonstrate this, 
we  evaluate the   correction to $S^{(1)}_{\rm sat}$ from the  second saddle  point for the $2A$ orbifold. 
We then  briefly discuss the implication of this constant for the geometric description of the dyons. 

The second property we study of low charge dyons is the Fourier-Jacobi coefficients
of the inverse Siegel modular form at fixed  magnetic charge of the dyon. 
The Fourier-Jacobi coefficients enumerate the degeneracy of dyons for arbitrary
electric charge and angular momentum but at fixed magnetic charge. We study the 
case of the first two non-trivial magnetic charges. 
We show that these Fourier-Jacobi coefficients are meromorphic Jacobi-forms which 
can be decomposed into an Appell-Lerch sum, the polar part and a finite 
term which is a mock modular form.  The finite part captures the degeneracy of single centered 
dyons. 
We perform this decomposition for all the orbifolds listed in table \ref{tt}. 
This decomposition was done in \cite{Dabholkar:2012nd} for the 
un orbifolded theory to  reasonably high magnetic charges. 
For the CHL orbifolds corresponding to $pA$ with $p = 2, 3, 5, 7$ at zero magnetic charge $P^2=0$
was done in the appendix A.5  of \cite{Bossard:2018rlt}. 
In this paper we have generalized these observations to all the other CHL orbifolds as well as the others 
in table \ref{tt} to the first non-trivial order $P^2 =2$. 
We observe that at this order there is a crucial identity among meromorphic Jacobi forms
\footnote{See equation (\ref{basicident}) for the identity.} that allows us to 
use the decompositions found by \cite{Dabholkar:2012nd}. In fact for all the orbifolds we show 
that the mock modular form that occurs at the level of $P^2=2$ is the 
generating function of Hurwitz-Kronecker class numbers found 
by \cite{Dabholkar:2012nd} for the un orbifolded theory.  
Thus our analysis generalizes these observations to all the orbifolds..

Finally we examine  the $1/4$ BPS dyons  in   orbifolds of type II compactifications  on $T^6$ which 
preserve ${\cal N}=4$ supersymmetry.  These theories were originally constructed in \cite{Sen:1995ff} and 
the partition function of dyons in these theories was obtained in  \cite{David:2006ru}. 
Our objective to study these models is to examine the  positivity conjecture of \cite{Sen:2010mz} on the Fourier coefficients
of the inverse Siegel modular forms which correspond to dyon partition functions. 
The crucial assumption that goes in to the positivity conjecture of \cite{Sen:2010mz} is that
if the charges of the dyon is  such that it is single centered, then the only sign to the index 
evaluating of the $1/4$ BPS state arises from the fermionic zero modes that arise due to the
breaking of ${\cal N}=4$ supersymmetry by the dyon. This is because the single centered dyon is assumed to 
be spherically symmetric and therefore carries zero angular momentum. 
We study the signs of the  Fourier coefficients that arise in the partition function of dyons 
of the $\mathbb{Z}_2$ and $\mathbb{Z}_3$ orbifolds of type II compactifications on $T^6$. 
We show that when the charges satisfy the condition that the dyon is single centered the sign of the 
index violates the positivity conjecture of \cite{Sen:2010mz}. 
We also use the criteria of subtracting out the polar part in the Fourier-Jacobi decomposition 
to identify the single dyons given a fixed magnetic charge and show that there 
are violations of the positivity conjecture. 
In fact we show that there is an infinite class of dyons which are single centered 
as well with magnetic charge $P^2 = 2$ which violate the positivity conjecture in the $\mathbb{Z}_2$
orbifold. 
These observations indicate the these dyons might posses hair modes which contribute to the sign 
of the $1/4$ BPS  index. This was one of the suggested ways, the positivity conjecture of 
\cite{Sen:2010mz} might be violated. 

The common thread which runs through  these results is that the all these properties 
are found at low charges of dyons  and in orbifold models of ${\cal N}=4$ compactifications. 
We see that there  is more to be learned about $1/4$ BPS dyons and their geometric 
description as black holes. 

The organization of the paper is as follows. 
In section \ref{sec2}, we compare the statistical entropy of low charge dyons and the exact degeneracies 
for all orbifolds models  given in table \ref{tt} and show that the constant $C_1$ contributes
crucially to the entropy to $S^{(1)}_{\rm stat}$. 
In section  \ref{sec3} we decompose the Fourier-Jacobi coefficient  that occur in the expansion 
of the inverse Siegel modular form for the first two non-trivial magnetic charges but 
arbitrary electric and magnetic charge in terms of an Appell-Lerch sum and a mock modular form. 
This is performed in detail for the $2A$ orbifold. 
In section \ref{sec4} we study the violation of the positivity conjecture by the Fourier coefficients
of the $\mathbb{Z}_2$ and $\mathbb{Z}_3$ toroidal orbifolds. 
Section \ref{sec5} contains our conclusions. 
Appendix \ref{appen1} contains the details regarding evaluating the constant $C_1$.

\section{Degeneracy  and statistical entropy at small  charges} \label{sec2}

Consider ${N=4}$ string compactifications obtained by considering 
type II  B theories on $K3\times T^2/\mathbb{Z}_N$ where the $Z_N$ acts 
as a automorphism  $g'$ on $K3$ together with a $1/N$ shift on one of the 
circles of $S^1$.  The action  $g'$  corresponds to the $26$ classes of 
the Mathieu group $M_{23}$.  For eg. The classes $pA$ with $p = 2, 3, 4, 5, 6, 7, 8$
are known as Nikulin's automorphism. These compactifications are also known
as CHL compactifications and they  were  introduced first as generalizations of 
models which dual to heterotic string compactifications with ${\cal N}=4$ 
supersymmetry \cite{Chaudhuri:1995ve,Chaudhuri:1995dj}.
Essentially these compactifications
reduce the rank of the gauge group but preserve ${\cal N}=4$ supersymmetry. 

These CHL compactifications  admit quarter  BPS dyons.  Let the charge vector for 
these dyons be $(\vec Q, \vec P)$ in the heterotic frame. Let $d(\vec Q, \vec P)$ 
denote  the  difference between the number of 
bosonic and fermions quarter BPS multiplets. In the region of 
moduli space where the type IIB theory is weakly coupled,  $d(\vec Q, \vec P)$ 
 is given by
\begin{eqnarray}\label{degen}
d(Q, P)  = \frac{1}{N} ( -1)^{ Q\cdot P +1} 
\int_{{\cal C}} d\tilde\rho d\tilde\sigma d \tilde v \; 
e^{-\pi i ( N \tilde\rho Q^2 + \tilde\sigma P^2/N + 2 \tilde v Q\cdot P ) }
\frac{1}{ \tilde \Phi_k ( \tilde \rho, \tilde \sigma , \tilde v) }.
\end{eqnarray}
The contour ${\cal C}$ is defined over a 3 dimensional subspace of the 
3 complex dimensional space $(\tilde \rho = \tilde\rho_1 + i \tilde\rho_2, \tilde\sigma = 
\tilde\sigma_1 + i \tilde\sigma_2,
\tilde v = \tilde v_1 + i \tilde v_2 ) $. 
\begin{eqnarray}\label{contour}
\tilde\rho_2 = M_1, \qquad \tilde\sigma_2 = M_2, \qquad \tilde v_2 = - M_3, \\ \nonumber
0\leq \tilde\rho_1 \leq 1, \qquad 0 \leq \tilde\sigma_1 \leq N, \qquad 0 \leq \tilde v_1 \leq 1.
\end{eqnarray}
Here $M_1, M_2, M_3$ are  positive numbers, which are fixed and large and
$M_3 << M_1, M_2$.   The contour essentially implies that we perform the expansions 
first in $e^{2\pi i \tilde \rho},  e^{2\pi i \tilde \sigma}$ and then perform the expansion in $e^{-2\pi i \tilde v}$. 
The function
$\tilde \Phi$, occurring in  integrand is a Siegel modular form transforming under 
a subgroup of $Sp(2, \mathbb{Z})$ with weight $k$. Explicitly, it is given by 
\begin{eqnarray}\label{siegform}
\tilde{\Phi}(\rho,\sigma,v)&=&e^{2\pi i(\tilde\rho+ \tilde\sigma/N+ \tilde v)}\\ \nn
&&\prod_{b=0,1}\prod_{r=0}^{N-1}
\prod_{\begin{smallmatrix}k'\in \mathbb{Z}+
	\frac{r}{N},l\in \mathcal{Z},\\ j\in 2\mathbb{Z}+b\\ k',l\geq0, \; j<0\;  k'=l=0\end{smallmatrix}}
(1-e^{2\pi i(k'\sigma+l\rho+jv)})^{\sum_{s=0}^{N-1}e^{2\pi isl/N}c_b^{r,s}(4k'l-j^2)}.
\end{eqnarray}
Here  $N$ is the order of the orbifold  $g'$.  The coefficients 
$c_b^{(r,s)}$ are read out from the expansion of the elliptic genus of $K3$ twisted 
by the action of $g'$.  
The twisted elliptic genus of $K3$ and its expansion is defined by 
\begin{eqnarray} \label{expelipg}
F^{(r, s)}(\tau, z) &=& \frac{1}{N} {\rm Tr}_{RR\; g^{\prime r } }
\left[ (-1)^{F_{K3} + \bar F_{K3} } 
g^{\prime s} e^{ 2\pi i z F_{\rm K3} } 
q^{L_0 - \frac{c}{24} } \bar q ^{\bar L_0 - \frac{\bar c}{24}} \right],  \\ \nonumber
&=& \sum_{b=0}^1 \sum_{j \in 2\mathbb{Z} + b, \; n \in \mathbb{Z}/N} 
c_b^{(r, s) }( 4n - j^2) e^{2\pi i n \tau + 2\pi i j z} .\\ \nonumber
& & \qquad\qquad \qquad 0 \leq r, s\leq N-1.
\end{eqnarray}
The trace in the above equation is taken over the Ramond-Ramond sector of the 
${\cal N}=(4, 4)$ super conformal field theory of $K3$ with central charge 
$(6, 6)$ and $F$ refers to the Fermion number. 
The twisted elliptic genera for the $g'$ belonging to all the conjugacy classes of 
$M_{23}\subset M_{24}$ has been evaluated in 
\cite{Gaberdiel:2012gf,Chattopadhyaya:2017ews}. 
They take the form 
\begin{eqnarray}\label{explielip}
F^{(0, 0)} ( \tau, z) &=& \alpha_{g'}^{(0, 0)} A( \tau, z) ,  \\ \nonumber
F^{(r, s) } ( \tau, z) &=& \alpha_{g'}^{(r, s) } A( \tau, z) + \beta_{g'}^{(r, s) }(\tau)  B(\tau, z) , 
\\ \nonumber
 && \qquad\qquad r, s \in \{0, 1, \cdots N-1 \} \; \hbox{with} ( r, s) \neq (0, 0), 
\end{eqnarray}
with 
\begin{eqnarray} \label{defab}
A (\tau, z) = \frac{\theta_2^2( \tau, z) }{ \theta_2^2 ( \tau, 0 )} 
+ \frac{\theta_3^2( \tau, z) }{ \theta_3^2 ( \tau, 0 )} 
+ \frac{\theta_4^2( \tau, z) }{ \theta_4^2 ( \tau, 0) } , 
\\ \qquad
B(\tau, z) = \frac{\theta_1^2(\tau, z) }{\eta^6(\tau) }.
\end{eqnarray}
$A(\tau, z)$ and $B(\tau, z) $ are Jacobi forms that transform under 
$SL(2, \mathbb{Z})$ with index 1 and weight 0 and $-2 $ respectively. 
The $\alpha_{g'}^{(r, s)}$ in (\ref{explielip}) are numerical constants and 
$\beta_{g'}^{(r, s) }$ are weight 2 modular forms which transform under 
$\Gamma_0(N)$.   For all $g'$ corresponding to the conjugacy classes 
listed in table \ref{tt}, the list of the twisted elliptic genera can be found in 
appendix E of \cite{Chattopadhyaya:2017ews}. 

\begin{table}[H]
	\renewcommand{\arraystretch}{0.5}
	\begin{center}
		\vspace{0.5cm}
				\begin{tabular}{|c|c|}
					\hline
					&  \\
					Conjugacy Class & Order \\
					\hline
					&  \\
					1A & 1\\
					2A & 2\\
					3A & 3\\
					5A & 5\\
					 7A & 7\\
					 11A & 11\\
					 23A/B & 23 \\
					\hline
					 & \\
					 4B & 4\\
					 6A & 6 \\
					 8A & 8 \\
					 14A/B & 14\\
					 15A/B & 15\\
					\hline
				\end{tabular}
		\end{center}
	\vspace{-0.5cm}
\caption{{\footnotesize{  
Conjugacy classes of $M_{24}$ studied in the paper.\\ 
We refer to the classes 2A, 3A, 5A, 7A, 4B, 6A, 8A  as the CHL orbifolds.}}} \label{tt}
\renewcommand{\arraystretch}{0.5}
\end{table}
\noindent
The weight $k$, of the Siegel modular form is given by 
\begin{equation}
k = \frac{1}{2} \sum_0^{N-1} c_0^{(0, s)} ( 0) .
\end{equation}
The weights of the Siegel modular forms corresponding to the twisted elliptic genera
constructed in this paper is listed in table \ref{t3}

\vspace{.4cm}
\begin{table}[H] 
\renewcommand{\arraystretch}{0.5}
\begin{center}
\vspace{-0.5cm}
\begin{tabular}{|c|c|c|c|c|c|c|}
\hline
 & & & & & & \\
Type 1 & pA & 4B & 6A & 8A & 14A & 15A \\
 & & & & & & \\
\hline
 & & & & & & \\
Weight  & $\frac{24}{p+1}-2$ & 3 & 2 & 1 & 0 & 0 \\
\hline
\end{tabular}
\end{center}
\vspace{-0.5cm}
\caption{Weight of Siegel modular forms corresponding to classes in $M_{23}$}
\label{t3}
\renewcommand{\arraystretch}{0.5}
\end{table}

Now using the twisted elliptic genera and product form for the Siegel modular
form in (\ref{siegform}) we can obtain $d(\vec Q, \vec P)$ as defined by  its 
Fourier expansion in (\ref{degen}) for low values of the charges. 
For sufficiently large values of charges, the integral in (\ref{degen}) 
can be performed  by evaluating the leading saddle point. 
The behaviour of the Siegel modular form $\tilde \Phi$ at the  saddle point 
is determined by another modular form $\hat \Phi$ which is 
related to by  a  $Sp(2, \mathbb{Z})$  transformation. 
Consider the transformation 
\begin{equation} \label{modtrans}
\tilde\rho=\frac{1}{N}\frac{1}{2v-\rho-\sigma},\quad \tilde\sigma={N}\frac{v^2-\rho\sigma}{2v-\rho-\sigma},\quad \tilde v=\frac{v-\rho}{2v-\rho-\sigma},
\end{equation}
with inverse
\begin{equation}
\rho = \frac{\tilde\rho\tilde\sigma -\tilde v^2}{N \tilde \rho}, 
\quad 
\sigma = \frac{\tilde\rho\tilde\sigma - ( \tilde v - 1)^2}{N \tilde \rho}, 
\quad 
v = \frac{\tilde\rho\tilde\sigma - \tilde v^2 + \tilde v }{N \tilde \rho}.
\end{equation}
The saddle point is located at $v=0$. 
This transformation results in the change of measure which is given by 
\begin{equation}
d\tilde\rho d\tilde\sigma d\tilde v = - ( 2v - \rho - \sigma)^{-3} d\rho d\sigma d v \;.
\end{equation}
Substituting this transformation we obtain 
\begin{eqnarray}
d(\vec Q, \vec P) &=& - \frac{1}{N}(-1)^{Q\cdot P + 1} 
\int_{\hat {\cal C}} d \rho d  \sigma d v ( 2v - \rho - \sigma)^{-3} \times  \\ \nonumber
&& \exp \left[ -i \pi \left\{ 
\frac{ v^2 - \rho\sigma}{2v - \rho -\sigma} P^2 + \frac{1}{ 2 v - \rho - \sigma} Q^2 
+ 2 \frac{ v- \rho}{2 v - \rho -\sigma} Q\cdot P \right\} \right]  \\ \nonumber
& & \times \frac{1}{\tilde\Phi_k( \tilde\rho, \tilde \sigma, \tilde v) }.
\end{eqnarray}
Here  the variables $(\tilde\rho, \tilde \sigma, \tilde v)$ are now thought of functions
of the variables $(\rho, \sigma, v)$. The contour is also correspondingly mapped. 
Now the Siegel modular forms constructed from the twisted elliptic genus 
of $K3$ corresponding satisfy the relation
 \begin{eqnarray}  \label{prop1}
\tilde \Phi_k( \tilde \rho, \tilde \sigma, \tilde v) = 
\tilde \Phi_k( \tilde \sigma/N, \tilde\rho N, \tilde v) .
\end{eqnarray}
As shown in \cite{David:2006yn}, 
this property results from the following equation satisfied by the twisted 
elliptic genera
\begin{equation}
\sum_{s=0}^{N-1} e^{-2\pi i ls/N} F^{(r, s)} (\tau, z) 
= \sum_{s=0}^{N-1} e^{-2\pi i  r s/N} F^{(l, s)} ( \tau, z) .
\end{equation}
In \cite{David:2006yn} it was 
verified that this property holds for all the orbifolds belonging to the 
class $pA$ with $p = 1, 2, 3, 5, 7$. We have verified that this property remains 
to be true for all the orbifolds $g'$ listed in table \ref{tt}. This includes all the CHL orbifolds 
in addition to the new ones in the conjugacy class of $M_{23}$. 
Now using (\ref{prop1}) with the transformation (\ref{modtrans}) we obtain 
\begin{eqnarray}\label{sp2trans}
\tilde 
\Phi_k( \frac{v^2 - \rho\sigma}{2v -\rho-\sigma}, \frac{1}{2v - \rho-\sigma},  \frac{v - \rho}{2v -\rho-\sigma}), \\ \nonumber
&=& -(i)^k C_1 ( 2 v - \rho - \sigma)^k \hat \Phi(\rho, \sigma, v) .
\end{eqnarray}
The last line defines the modular function $\hat \Phi_k$ related to 
$\tilde\Phi_k$ by the $Sp(2, \mathbb{Z})$ transformation. 
For the case of the unorbifolded  $K3$, $k =10$ 
and since the Igusa cusp form is unique $\hat \Phi_{10}$ coincides with 
$\tilde\Phi_{10}$ and the constant $C_1=1$. 
Thus the leading saddle point is determined by the behaviour of the 
new modular form $\hat \Phi_k$ at $v\rightarrow 0$. 
Here $C_1$ is constant which is non-trivial for all the orbifolds in table \ref{tt} 
and  we will show it plays an important role \footnote{ As we will subsequently 
demonstrate, we have chosen the 
phases so that $C_1$ is real.}. 
It can be shown using the product representation 
of $\hat\Phi_k$  that at $v\rightarrow 0$
\begin{equation}\label{factorhatphi}
\hat\Phi_k(\rho, \sigma,  v)|_{\rm v\rightarrow 0}  = -4\pi^2 v^2 g(\rho) g(\sigma)
\end{equation}
where $g(\tau)$ is a specific $\Gamma_0(N)$ form for each of the orbifold $g'$ of 
weight $ k+2$
For example for the $2A$ orbifold 
\begin{equation}
g(\tau) = \eta^8(\tau) \eta^8(2\tau)
\end{equation}
The list of the function $g(\tau)$ for each of the orbifolds is given in  table  \ref{t6}.
This was obtained for $pA,  p =2, 3, 5, 7$ orbifolds in \cite{David:2006ji} and 
for the remaining  orbifolds  of  table \ref{t6}  in \cite{Chattopadhyaya:2017ews}
\begin{table}[H]
\renewcommand{\arraystretch}{0.5}
\begin{center}
\vspace{0.5cm}
\begin{tabular}{|c|c|}
\hline
 & \\
Conjugacy Class  & $g^{(k+2)} (\rho)$ \\ \hline
 &   \\
$pA$ & $\eta^{k+2}(\rho)\eta^{k+2}(p\rho)$\\
 & \\
4B & $\eta^4(4\rho)\eta^2(2\rho)\eta^{4}(\rho)$\\
 & \\
6A & $\eta^2(\rho)\eta^2(2\rho)\eta^2(3\rho)\eta^2(6\rho)$\\
 & \\
8A & $\eta^2(\rho)\eta(2\rho)\eta(4\rho)\eta^2(8\rho)$\\
 & \\
14A &  $\eta(\rho)\eta(2\rho)\eta(7\rho)\eta(14\rho)$ \\
 & \\
15A &  $\eta(\rho)\eta(3\rho)\eta(5\rho)\eta(15\rho)$\\
  & \\
\hline
\end{tabular}
\end{center}
\vspace{-0.5cm}
\caption{Factorization of $\hat \Phi_k(\rho, \sigma, v) $ as $\lim v\rightarrow 0$ , $p\in \{1,2,3,5,7,11\}$} \label{t6}
\renewcommand{\arraystretch}{0.5}
\end{table}

The end result of evaluating the saddle at 
$v=0$ and the one loop determinant is the following. 
Let us define the statistical entropy by 
\begin{equation}
S_{\rm stat} \equiv \ln d(\vec Q, \vec P) 
\end{equation}
To obtain the result of $S_{\rm stat}$ we need to consider the statistical entropy  function 
\begin{equation} \label{statent}
S(\tau) = \frac{\pi}{2\tau_2} | Q - \tau P|^2 - \ln g(\tau) - \ln g(-\bar \tau) 
- (k+2) \ln (2\tau_2) -  \ln ( NC_1) + O( Q^{-2}, P^{-2}) 
\end{equation}
where $\tau = \tau_1 + i \tau_2$. 
Then the statistical entropy
 is obtained by evaluating $S(\tau)$  at its extremum $\tau_{\rm{extremum}}$
\begin{equation}
S_{\rm stat}^{(1)}  = S(\tau)|_{\tau_{\rm{extremum}}}.
\end{equation}
The superscript $(1)$ refers to the fact that the statistical entropy is 
obtained at one loop from the leading saddle. 
Note the presence of the constant $C_1$ in the statistical entropy function
(\ref{statent}), this is the
constant that relates the  generating function for the degeneracies 
$\tilde\Phi_k$ and its $Sp(2, \mathbb{Z})$ transform  $\hat\Phi_k$ 
as given in (\ref{sp2trans}). 
As mentioned earlier,  $C_1 =1$ for the unorbifolded $K3\times T^2$ compactification, 
and therefore there is 
no contribution from this term to the statistical entropy function. 

Our goal in the rest of this section is to evaluate this constant 
and compare its contribution in $S_{\rm stat}^{(1)}$ 
 to the exact entropy $S_{\rm stat} = \ln d(\vec Q, \vec P)$. 
 We will perform this comparison  
 for low values of charges as it is clear that for very large values of 
 charges this constant will not play a relevant role. 
 Such a comparison  for low values of charges 
 of the one loop statistical entropy function
 with the exact entropy was made for the un-orbifolded  $K3\times T^2$ compactification
 in \cite{Sen:2007qy} and later in \cite{Banerjee:2008ky}. 
 It was seen that the statistical entropy function 
 at one loop agrees with the exact entropy to $2\%$ even for the lowest 
 admissible charge. 
 We will extend this comparison for all the orbifolds  $g'$ listed in table \ref{tt}. 
 We will see that the constant $C_1$ is non-trivial and depends on the orbifold
 and contributes crucially towards $S_{\rm stat}^{(1)}$

\subsection{$\hat \Phi_k$ and the  constant $C_1$} 

In this section we determine the constant $C_1$ which occurs in the 
modular transformation relating $\tilde\Phi_k$ and 
$\hat \Phi_k$ 
 (\ref{sp2trans}). 
 We will first follow the first principle method which defines $\hat \Phi_k$ in terms of the 
 a `threshold integral' relating it to the form  $\tilde\Phi_k$ and obtain $C_1$. 
 We then perform a  simple cross check, by using the  factorization property 
 of $\hat\Phi_k$ given in (\ref{factorhatphi}). 
 Before we proceed we  simplify the modular transform 
 relating  these Siegel modular  forms. 
 We can also write (\ref{sp2trans}) as 
 \begin{equation}\label{sp2trans1}
 \tilde\Phi_k( \tilde\rho', \tilde\sigma' , \tilde v') =  -( i)^k C_1 ( \tilde \sigma')^{-k} 
 \hat \Phi_k\left( \tilde \rho' - \frac{ (\tilde v' )^2}{\tilde\sigma'}, 
 \tilde\rho' - \frac{(\tilde v' -1)^2}{\tilde\sigma'}, \tilde\rho' - \frac{\tilde v'^2}{\tilde\sigma'} 
 + \frac{\tilde v'}{\tilde\sigma'} \right)
 \end{equation}
 Here we have defined
 \begin{equation}
 \tilde \rho'  = \frac{v^2 - \rho\sigma}{ 2 v - \rho - \sigma} , 
 \quad \tilde\sigma' = \frac{1}{2 v - \rho -\sigma} , 
 \quad
 \tilde v' = \frac{v - \rho}{ 2 v - \rho -\sigma}
 \end{equation}
 and its inverse
 \begin{equation}
 \rho = \frac{\tilde\rho ' \tilde\sigma ' - \tilde v^{\prime 2}}{\tilde\sigma'}, 
 \quad 
 \sigma = \frac{\tilde\rho'\tilde\sigma' - (\tilde v' - 1)^2}{\tilde \sigma'}, 
 \quad
 \tilde v =  \frac{\rho'\tilde\sigma' - \tilde v^{\prime 2} + \tilde v}{\tilde \sigma'}.
 \end{equation}
 $\hat\Phi_k$ is invariant under the transformation \cite{Jatkar:2005bh}
 \footnote{This fact will also subsequently be evident from the final result 
 of $\hat\Phi_k$ in terms of a `threshold integral'.}. 
 \begin{equation}\label{invhat}
 \hat\Phi_k( \rho, \sigma, v) = \hat\Phi_k( \rho, \sigma + \rho - 2 v, v - \rho).
 \end{equation}
 Using this invariance we can re write the modular transformation 
 (\ref{sp2trans1}) as
 \begin{equation} \label{sp2trans2}
  \tilde\Phi_k( \tilde\rho', \tilde\sigma' , \tilde v')  = - (i)^{k} C_1 (\tilde\sigma')^{-k} 
  \hat\Phi_k\left( \tilde{\rho'} - \frac{\tilde v^{\prime 2} }{\tilde\sigma'},
  - \frac{1}{\tilde\sigma'}, \frac{\tilde v'}{\tilde\sigma} \right) .
 \end{equation}
 To avoid cluttering, we will now refer to $(\tilde\rho', \tilde\sigma', \tilde v')$  as
 $( \tilde\rho, \tilde\sigma, \tilde v )$. 
 The constant $C_1$ can be found by  examining the construction of Siegel 
 modular forms $\tilde\Phi_k$ and $\hat\Phi_k$ using `threshold integrals'. 
 For $\tilde\Phi_k$ we consider  the integral
 \begin{eqnarray}\label{threshold1}
 \tilde {\cal I}(\tilde\rho, \tilde\sigma, \tilde v )  &=& \sum_{r, s =0}^{N-1}\sum_{b= 0}^1
 \tilde {\cal I }_{r, s,  b } \\ \nonumber
 \tilde{\cal I}_{r, s, b } ( \tilde\rho, \tilde\sigma, \tilde v) 
 &=& \int_{\cal F} \frac{d^2\tau}{\tau_2} \sum_{m_1, m_2, n_2\in\mathbb{Z}, n_1\in 
 \mathbb{Z} + \frac{r}{N}, j\in 2\mathbb{Z} + b } 
 \exp\left[ 2\pi i \tau ( m_1n_1 + m_2 n_2 + \frac{j^2}{4} )\right] \times \\ \nonumber
 & & \exp\left( - \frac{\pi \tau_2}{\tilde Y} | n_2 ( \tilde\rho \tilde \sigma - \tilde v^2 ) + j \tilde v
 + n_1 \tilde\sigma  -m_1 \tilde\rho +  m_2 |^2 \right) e^{2\pi i m_1 s /N} h_{b}^{(r, s)}(\tau) 
 \\ 
 & & \qquad\qquad 0\leq r, s \leq (N-1) 
 \end{eqnarray}
 where 
 \begin{equation}
\tilde Y=\det{\rm Im } ( \tilde \Omega), \qquad \tilde\Omega  =
 \left(\begin{matrix}
\tilde\rho & \tilde v\\ \tilde v  & \tilde\sigma
\end{matrix}\right)
\end{equation}
and  $h_b^{(r, s)}$ are found by expanding the twisted elliptic genus as
 \begin{equation}
 F^{(r,s)}(\tau, z) =h_1^{r,s}(\tau)\theta_2(2\tau,2z)+h_0^{r,s}\theta_3(\tau)(2\tau,2z).
 \end{equation}
 We can use the method of orbits to evaluate this integral as done in 
 \cite{David:2006ji} and we obtain 
 \begin{eqnarray} \label{tildint}
 \tilde{\cal I} ( \tilde\rho, \tilde\sigma, \tilde v) 
 &=& - 2 \ln [ {\rm det }{ \rm Im} \tilde \Omega) ^k ] - 2 \ln 
 \tilde \Phi_k( \tilde\rho, \tilde\sigma, \tilde v)  -  2 \ln 
\bar{ \tilde \Phi}_k( \tilde\rho, \tilde\sigma, \tilde v)  - 2 k \ln\kappa,  \nonumber\\
& &\kappa = \frac{8\pi}{3 \sqrt{3} } e^{ 1- \gamma_E} ,
\end{eqnarray}
where $ \gamma_E$  is the Euler-Mascheroni constant. 

Now lets go over to the modular form $\hat\Phi_k$.  From the  $Sp(2, \mathbb{Z})$
transformation given in (\ref{sp2trans2}) we see that this is equivalent to 
\begin{equation} \label{sp2trans3}
\tilde\Phi( \rho - \frac{v^2}{\sigma}, - \frac{1}{\sigma}, - \frac{v}{\sigma} ) = 
- (-i)^k C_1 (-\sigma)^{k} \hat\Phi_k( \rho, \sigma, v) .
\end{equation}
From this transformation and from the threshold integral in (\ref{threshold1}) we see 
that we can obtain $\hat\Phi_k$ by the following replacements
\begin{equation}
m_2 \rightarrow n_1, \quad, n_1 \rightarrow -m_2, \quad m_1 \rightarrow -n_2, 
\quad n_2 \rightarrow m_1.
\end{equation}
Therefore  to construct $\hat\Phi$ we can consider the integral
\begin{eqnarray}\label{thresholdhat}
\hat{\cal I} (\rho, \sigma, v) &=& \sum_{r, s=0}^{N-1} \hat {\cal I}(\rho, \sigma, v) , \\ \nonumber
\hat{\cal I}(\rho, \sigma, v) 
&=& \int_{\cal F} \frac{d^2\tau}{\tau_2} \sum_{m_1, n_1, n_2\in\mathbb{Z}, m_1\in 
 \mathbb{Z} - \frac{r}{N}, j\in 2\mathbb{Z} + b } 
 \exp\left[ 2\pi i \tau ( m_1n_1 + m_2 n_2 + \frac{j^2}{4} )\right] \times \\ \nonumber
 & & \exp\left( - \frac{\pi \tau_2}{ Y} | n_2 (  \rho  \sigma - v^2 ) + j \ v
 + n_1 \sigma  -m_1  \rho +  m_2 |^2 \right) e^{-2\pi i n_2 s /N} h_{b}^{(r, s)}(\tau) 
 \\  \nonumber
 & & \qquad\qquad 0\leq r, s \leq (N-1)  .
 \end{eqnarray}
Examining the integrals and using the relation between the coordinates we can 
see that
\begin{equation}\label{equality}
\hat{\cal I} ( \rho, \sigma, v) = \tilde{\cal I} ( \tilde\rho, \tilde\sigma, \tilde v).
\end{equation}
In fact  using a similar analysis one can also verify the invariance of 
 of $\hat\Phi$ under the transformation  given 
in (\ref{invhat}). 

Evaluating the integral through the method of orbits we obtain 
\begin{eqnarray} \label{reshatint}
\hat {\cal I} (\rho, \sigma, v)  &=& 
- 2 \ln [ {\rm det }{ \rm Im}  \Omega) ^k ] - 2 \ln 
 \hat \Phi_k( \rho, \sigma, \tilde v)  -  2 \ln 
\bar{ \hat \Phi}_k( \rho, \sigma,  v)  - 2 k \ln\kappa \\ \nonumber
& & - 2 (k+ 2) \ln(N),
\end{eqnarray}
where 
\begin{eqnarray}\nn
\hat{\Phi}(\rho,\sigma,v)&=&-e^{2\pi i(\rho+\sigma +v)}\prod_{r,s=0}^{N-1}\prod_{\begin{smallmatrix}
	k',l,b\in \mathbb{Z}\\k',l,j>0\end{smallmatrix}} \left(1-e^{2\pi i r/N}\exp(2\pi i(k'\sigma+l\rho+jv))\right)^{\frac{1}{2}c^{r,s}(4k'l-j^2)}\\
&& \qquad\qquad\qquad \prod_{r,s=0}^{N-1}\prod_{\begin{smallmatrix}
	k',l,b\in \mathbb{Z}\\k',l,j>0\end{smallmatrix}} 
 \left(1-e^{-2\pi i r/N}\exp(2\pi i(k'\sigma+l\rho+jv))\right)^{\frac{1}{2}c^{r,s}(4k'l-j^2)}.
 \nonumber \\
\end{eqnarray}
The details of the evaluation of the threshold integral again  follow the methods of
 \cite{David:2006ji}. 
However since we are interested in keeping track of the constant 
in the last line of (\ref{reshatint}) we give some of the details of this in the appendix \ref{appen1}

Using the equality (\ref{equality}) and substituting  the modular transformation 
(\ref{sp2trans3}) in 
equations
(\ref{tildint}) and (\ref{reshatint}) we find that 
\begin{equation}\label{resc1}
C_1 =  N^{\frac{k+2}{2} }
\end{equation}
It is clear that this approach does not fix the phase we have chosen  in the 
modular transformation (\ref{sp2trans}). 
To  fix this  phase and  also perform a cross check on $C_1$, let us 
examine  the relation given (\ref{sp2trans}) in the $\tilde v\rightarrow 0$ limit. 
From the product representation of $\tilde \Phi_k$ given in (\ref{siegform}) we 
can show that  $\tilde\Phi_k$ factorizes as
\begin{equation} \label{factildphi}
\lim_{\tilde v \rightarrow 0} 
\tilde\Phi_k(\tilde \rho, \tilde\sigma, \tilde v) 
= - 4\pi^2 g(\tilde \rho ) h( \tilde \sigma) 
 \end{equation}
 where $g(\tilde \rho)$ and $h(\tilde \sigma)$ are modular forms of weight $k+2$ transforming 
 under subgroups of $\Gamma_0(N)$. 
 The list of these forms for each  the $\tilde \Phi_k$  corresponding to the 
orbifolds considered in this paper is given in table \ref{t5}.

\begin{table}[H]
\renewcommand{\arraystretch}{0.5}
\begin{center}
\vspace{0.5cm}
\begin{tabular}{|c|c|c|c|}
\hline
 & & & \\
Conjugacy Class & $k$ & $g  (\rho)$ & $h(\sigma)$\\ \hline
 & & & \\
$p$A & $\frac{24}{p+1}-2$ &$\eta^{k+2}(\rho)\eta^{k+2}(p\rho)$  & $\eta^{k+2}(\sigma)\eta^{k+2}(\sigma/p)$\\
 & & & \\
4B & 3 & $\eta^4(4\rho)\eta^2(2\rho)\eta^{4}(\rho)$ &  $\eta^{4}(\frac{\sigma}{4})\eta^{2}(\frac{\sigma}{2})\eta^{4}(\sigma)$\\
 & & & \\
6A & 2& $\eta^2(\rho)\eta^2(2\rho)\eta^2(3\rho)\eta^2(6\rho)$ & $\eta^2(\sigma)\eta^2(\frac{\sigma}{2}) \eta^2(\frac{\sigma}{3})\eta^2(\frac{\sigma}{6})$\\
 & & & \\
8A & 1 &  $\eta^2(\rho)\eta(2\rho)\eta(4\rho)\eta^2(8\rho)$ & $\eta^2(\sigma)\eta(\frac{\sigma}{2}) \eta(\frac{\sigma}{4})\eta^2(\frac{\sigma}{8})$\\
 & & & \\
14A & 0 &  $\eta(\rho)\eta(2\rho)\eta(7\rho)\eta(14\rho)$ &  $\eta(\sigma)\eta(\frac{\sigma}{2}) \eta(\frac{\sigma}{7})\eta(\frac{\sigma}{14})$\\
 & & &\\
15A & 0 & $\eta(\rho)\eta(3\rho)\eta(5\rho)\eta(15\rho)$ &  $\eta(\sigma)\eta(\frac{\sigma}{3}) \eta(\frac{\sigma}{5})\eta(\frac{\sigma}{15})$\\
 & & & \\
\hline
\end{tabular}
\end{center}
\vspace{-0.5cm}
\caption{Factorization of $\tilde \Phi_k(\rho, \sigma, v) $ as $\lim v\rightarrow 0$ as shown in 
, $p\in \{1,2,3,5,7,11\}$} \label{t5}
\renewcommand{\arraystretch}{0.5}
\end{table}

Taking the limit $\tilde v\rightarrow 0$ in (\ref{sp2trans}), we obtain 
\begin{equation}\label{vlimit}
4\pi^2 g( \tilde\rho) h (\tilde \sigma) = -( i )^k C_1 \tilde\sigma^{-k}
( \frac{\tilde v}{\tilde\sigma} )^2 g( \tilde\rho) g( - \frac{1}{\tilde\sigma} ) 
\end{equation}
From the list of the weight $k+2$ forms $g$  given in  table \ref{t6} we see 
that   all $g$ obey the property
\begin{equation} \label{gtrans}
g(-\frac{1}{\tilde \sigma} ) = - ( -i)^k N^{ - (\frac{k+2}{2} )} h(\tilde\sigma)
\end{equation}
Now from substituting (\ref{gtrans}) into (\ref{vlimit}) we  confirm that 
$C_1$ is given by (\ref{resc1}).

\subsection{Comparison with   statistical entropy at one loop}\label{sec13}

We   compare  the logarithm of the  degeneracy  obtained
from the Fourier expansion given in (\ref{degen}) 
\begin{equation}
S_{\rm stat} = \ln d( Q, P) 
\end{equation}
and the statistical entropy  at one loop  which is given by 
\begin{eqnarray}\label{statisent1}
S_{\rm stat}^{(1)} &=&  
 \frac{\pi}{2\tau_2} | Q - \tau P|^2 - \ln g(\tau) - \ln g(-\bar \tau) 
- (k+2) \ln (2\tau_2) -  \ln ( NC_1) ,  \\ \nonumber
\tau_1 &=& \frac{Q\cdot P}{P^2},  \qquad  \tau_2 = \frac{1}{P^2} \sqrt{ Q^2 P^2 - (Q\cdot P)^2}, 
\qquad
C_1 = N^{\frac{k+2}{2}}.
\end{eqnarray}
We perform this comparison 
for all the orbifolds $g'$ listed in table \ref{tt} for low value of charges. 
It is easy to evaluate the logarithm of the degeneracy by performing 
the Fourier expansion in Mathematica for low value of charges. 
Then the comparison with the statistical entropy  is done both 
with and without constant $-\ln(NC_1)$. 
We see that for low values of charges, the constant is crucial 
in bringing the agreement of the statistical entropy to within 
a few percent of the actual degeneracy. 

In the tables below $\delta$ and $ \delta' $  are defined as follows
\begin{eqnarray}
\delta &=& \frac{  S_{\rm stat}- S^{(1)}_{\rm stat} }{ S_{\rm stat}} \times 100 ,  \\ \nonumber
\delta' &=&  \frac{ \{S^{(1)} _{\rm stat}+ \ln(NC_1) \} - S_{\rm stat} }{ S_{\rm stat}} \times 100
\end{eqnarray}
Thus $\delta'$ measures the percentage difference from $S_{\rm stat}$ 
 without the constant $-\ln(NC_1)$
in the statistical entropy at one loop. 
We list the  comparisons for orbifolds of the order $N$ where $N$ is prime first and then 
move to the non-prime cases. 

\begin{table}[H]
	\renewcommand{\arraystretch}{0.5}
	\begin{center}
		\vspace{0.5cm}
		\begin{tabular}{|c|c|c|c|c|c|}
			\hline
			& & & & & \\
			$(Q^2,\,P^2\, Q.P)$ & $d(Q, P )$ & $S_{\rm stat}$ & $S^{(1)}_{\rm stat}$ 
			& $\delta$ & $\delta '$ \\ 
			& & & & & \\
			\hline
			& & & & &  \\
			(1, 2, 0) & 2164 & 7.67971 & 7.28409 & 5.15 & 50.28  \\
			(1, 2, 1) & 360 & 5.8861 & 5.34077 & 9.26 & 68.14 \\
			(1, 4, 1) & 4352 & 8.37839 & 8.39542 & $-$0.2 & 41.16 \\
			(2, 4, 0) & 198144 & 12.1967 & 11.727 & 3.85 & 32.27 \\
			(1, 6, 1) & 36024 & 10.4919 & 11.1568 & $-$6.33 & 26.7 \\
			(3, 6, 0) & 15219528 & 16.5381 & 16.1699 &2.22 & 23.18 \\
			(3, 6, 3) & 149226 & 11.9132 & 11.624 & 2.43 & 31.52 \\
			(3, 6, 4) & 2164 &  7.67971 & 7.28409 & 5.15 & 50.28 \\
			\hline
		\end{tabular}
	\end{center}
	\vspace{0.5cm}
	\caption{ Comparison of the statistical entropy and statistical 
	entropy at one loop  for 2A orbifold}
	\renewcommand{\arraystretch}{0.5} \label{tt2a}
\end{table}

\begin{table}[H]
	\renewcommand{\arraystretch}{0.5}
	\begin{center}
		\vspace{0.5cm}
		\begin{tabular}{|c|c|c|c|c|c|}
			\hline
			& & & & & \\
			$(Q^2,\,P^2\, Q.P)$ & $d(Q, P) $ & $S_{\rm stat}$ & $S^{(1)}_{\rm stat}$ 
			& $\delta$ & $\delta'$ \\ 
			& & & & &  \\
			\hline
			& & & & & \\
			(2/3, 2, 0) & 540 & 6.29157 & 5.95751 & 5.31 & 75.16 \\
			(2/3, 4, 0) & 3294 & 8.09986 & 8.12528 & $-0.31$ &53.94  \\
			(2/3, 4, 1) & 378 & 5.93489 & 6.0378 & $-1.73$ & 72.31 \\
			(2/3, 6, 0)  & 16200 & 9.69277 & 10.3187 & $-6.46$ & 38.88 \\
			(2/3, 6, 1) & 2646 & 7.8808 & 8.58224 & $-8.90$ & 46.86 \\
			(4/3, 6, 0) & 128706 & 11.7653 & 11.4413 & 2.75 & 40.10 \\
			(4/3, 6, 1) & 37422 & 10.53 & 10.546 & $-0.15$ & 41.58 \\
			(2, 6, 0) & 820404 & 13.6176 & 13.2127 & 2.97 & 35.24 \\
			(2, 6, 1) & 318267 & 12.6706 & 12.546 & 0.98 & 35.66 \\
			(2, 6, 2) & 37818 & 10.5405 & 10.4379 & 0.97 & 42.66\\
			\hline
		\end{tabular}
	\end{center}
	\vspace{0.5cm}
	\caption{ Comparison of the statistical entropy and statistical 
	entropy at one loop  for
	  3A orbifold}
	\renewcommand{\arraystretch}{0.5}\label{tt3a}
\end{table}

\begin{table}[H]
	\renewcommand{\arraystretch}{0.5}
	\begin{center}
		\vspace{0.5cm}
		\begin{tabular}{|c|c|c|c|c|c|}
			\hline
			& & & & & \\
			$(Q^2,\,P^2\, Q.P)$ & $d(Q, P)$ & $S_{\rm stat}$ & $S^{(1)}_{\rm stat}$ & $\delta$ & $\delta'$ \\ 
			& & & & & \\
			\hline
			& & & & & \\
			(2/5, 2, 0) & 100 & 4.60517 & 4.57546 & 0.64 & 105\\
			(2/5, 4, 0) & 460 & 6.13123 & 6.30791 & $-2.88$ & 75.86\\
			(2/5, 4, 1) & 20 & 2.99573 & 2.87779 &3.94 & 165 \\
			(2/5, 6, 0)  & 1720 & 7.45008 & 8.08384 & $-$8.51 & 56.30  \\
			(2/5, 6, 1) & 125 & 4.82831 & 5.46281 & $-13.14$ & 86.85 \\
			(4/5, 6, 0) & 9180 & 9.12478 & 8.84163 & 3.10 & 56.02 \\
			(4/5, 6, 1) & 1460 & 7.28619 & 7.49625 & $-2.88$ & 63.38 \\
			(6/5, 6, 0) & 39960 & 10.5956 & 10.1953 & 3.78 & 49.35  \\
			(6/5, 6, 1) & 9345 & 9.1426 & 9.21235 & $-0.76$ & 52.05 \\
			(6/5, 6, 2) & 390 & 5.96615 & 5.88441 & 1.37 & 82.3 \\
			\hline
		\end{tabular}
	\end{center}
	\vspace{0.5cm}
	\caption{ Comparison of the statistical entropy and statistical 
	entropy at one loop  for
	5A orbifold}
	\renewcommand{\arraystretch}{0.5} \label{tt5a}
\end{table}

\begin{table}[H]
	\renewcommand{\arraystretch}{0.5}
	\begin{center}
		\vspace{0.5cm}
		\begin{tabular}{|c|c|c|c|c|c|}
			\hline
			& & & & & \\
			$(Q^2,\,P^2\, Q.P)$ & $d(Q, P)$ & $S_{\rm stat}$ & $S^{(1)}_{\rm stat}$ & $\delta$ & $\delta'$\\ 
			& & & & & \\
			\hline
			& & & & &  \\
			(2/7, 2, 0) & 36 & 3.58352 & 3.74248 & $-4.43$ & 131\\
			(2/7, 4, 0) & 138 & 4.92725 & 5.23041 & $-6.15$ & 92.58 \\
			(2/7, 6, 0)  & 444 & 6.09582 & 6.76722 & $-11.01$ & 68.79  \\
			(2/7, 6, 1) & 18 & 2.89037 & 3.22552 & $-11.59$ & 156.7 \\
			(4/7, 6, 0) & 1916 & 7.55799 & 7.35616 & 2.67 & 67.04 \\
			(4/7, 6, 1) & 210 & 5.34711 & 5.60863 & $-4.89$ & 86.09\\
			(6/7, 6, 0) & 6892 & 8.83812 & 8.49212 & 3.91 & 58.96 \\
			(6/7, 6, 1) & 1152 & 7.04925 & 7.23465 & $-2.63$ & 66.38 \\
			(6/7, 6, 2) & 18 & 2.89037 & 2.58629 & 10.52 & 178.83 \\
			\hline
		\end{tabular}
	\end{center}
	\vspace{0.5cm}
	\caption{Comparison of the statistical entropy and statistical 
	entropy at one loop for 7A orbifold.}
	\renewcommand{\arraystretch}{0.5}
\end{table}

\begin{table}[H]
	\renewcommand{\arraystretch}{0.5}
	\begin{center}
		\vspace{0.5cm}
		\begin{tabular}{|c|c|c|c|c|c|}
			\hline
			& & & & & \\
			$(Q^2,\,P^2\, Q.P)$ & $d(Q, P)$ & $S_{\rm stat}$ & $S^{(1)}_{\rm stat}$ & $\delta$ & $\delta'$ \\ 
			& & & & & \\
			\hline
			& & & & &  \\
			(6/11, 10, 0) & 4962 & 8.50956 & 8.32923 & 2.12 & 54.2\\
			(6/11, 10, 1) & 937 & 6.84268 & 7.13426 &$-4.26$ & 74.35  \\
			(6/11, 12, 0) & 11132 & 9.31758 & 9.22308 & 1.01 & 50.4 \\
			(6/11, 12, 1) & 2558 & 7.72223 & 8.16335 & $-5.7$ & 67.8 \\
			(6/11, 12, 2) & 72 & 4.27667 & 4.36939 & $-2.16$ & 114 \\
			(6/11, 22, 0) & 366378 & 12.8114 & 13.1652 & $-2.76$ & 40 \\
			(6/11, 22, 1) & 139955 & 11.8491 & 12.4283  & $4.88$ & 45.6 \\
			(6/11, 22, 2) & 12760 & 9.45407 & 10.0209 & $5.99$ & 56.7 \\
			(6/11, 22, 3) & 114 & 4.7362 & 4.86058 & $2.6$ & 103 \\
			\hline
		\end{tabular}
	\end{center}
	\vspace{0.5cm}
	\caption{Comparison of the statistical entropy and statistical 
	entropy at one loop for 11A orbifold.}
	\renewcommand{\arraystretch}{0.5} \label{tt11a}
\end{table}

\begin{table}[H]
	\renewcommand{\arraystretch}{0.5}
	\begin{center}
		\vspace{0.5cm}
		\begin{tabular}{|c|c|c|c|c|c|}
			\hline
			& & & & & \\
			$(Q^2,\,P^2\, Q.P)$ & $d(Q, P)$ & $S_{\rm stat}$ & $S^{(1)}_{\rm stat}$ & $\delta$ & $\delta'$ \\ 
			& & & & & \\
			\hline
			& & & & & \\
			(4/23, 8, 0) & 91 & 4.51086 & 3.63606   & 19.3 & 84.87  \\
			(6/23, 6, 0) & 103 & 4.63473 & 3.5562 & 23.27 & 78.2 \\
			(6/23, 8, 0) & 190 & 5.24702 & 4.21628 & 19.6 & 70 \\
			(6/23, 10, 0) & 312 & 5.743 & 4.86463 & 15.2 & 66.6 \\
			(6/23, 10, 1) & 74 & 4.30407 & 2.90311 & 32.5 &  76.7 \\
			\hline
		\end{tabular}
	\end{center}
	\vspace{0.5cm}
	\caption{Comparison of the statistical entropy and statistical 
	entropy at one loop for 23A orbifold.} \label{23a}
	\renewcommand{\arraystretch}{0.5}
\end{table}

\begin{table}[H]
	\renewcommand{\arraystretch}{0.5}
	\begin{center}
		\vspace{0.5cm}
		\begin{tabular}{|c|c|c|c|c|c|}
			\hline
			& & & & & \\
			$(Q^2,\,P^2\, Q.P)$ & $d(Q, P)$ & $S_{\rm stat}$ & $S^{(1)}$ & $\delta$ & $\delta'$\\  
			& & & & &  \\
			\hline
			& & & & & \\
			(1/2, 2, 0) & 176 & 5.17048 & 4.9493 & 4.28 & 98.12  \\
			(1/2, 4, 0) & 896 & 6.79794 & 6.84008 & $-$0.62 & 70.75 \\
			(1/2, 4, 1) & 80 & 4.38203 & 4.25615 & 2.87 & 113  \\
			(1/2, 6, 0) & 3616 & 8.19312 & 8.7606 &$-6.92$& 52.29 \\
			(1/2, 6, 1) & 480 & 6.17379 & 6.65185 & $-7.74$ & 70.85 \\
			(1, 4, 0) & 5024 & 8.52198 & 8.09089 & 5.06 & 61.99 \\
			(1, 4, 1) & 832 & 6.72383 &  6.68116 & 0.63 & 72.79 \\
			(3/2, 8, 0) & 491920 & 13.1061 & 12.7909 & 2.40 & 39.43 \\
			(3/2, 8, 1) & 196960 & 12.1908 & 12.1281 & 0.51 & 40.31 \\
			(3/2, 8, 2) & 23616 &  10.0697 & 10.0251 & 0.44 & 48.64\\
			\hline
		\end{tabular}
	\end{center}
	\vspace{0.5cm}
	\caption{Comparison of the statistical entropy and statistical 
	entropy at one loop for 4B orbifold.} \label{tt4b}
	\renewcommand{\arraystretch}{0.5}
\end{table}

\begin{table}[H]
	\renewcommand{\arraystretch}{0.5}
	\begin{center}
		\vspace{0.5cm}
		\begin{tabular}{|c|c|c|c|c|c|}
			\hline
			& & & & & \\
			$(Q^2,\,P^2\, Q.P)$ & $d(Q, P)$ & $S_{\rm stat}$ & $S^{(1)}_{\rm stat}$ & $\delta$ & $\delta'$ \\ 
			& & & & & \\
			\hline
			& & & & & \\
			(1/3, 2, 0) & 40 & 3.68888 & 3.50247 & 5.05 & 150 \\
			(1/3, 4, 0) & 148 & 4.99721 & 5.04572 & $-0.97$ & 106 \\
			(1/3, 6, 0) & 478 & 6.16961 & 6.61552 & $-7.23$ & 79.9 \\
			(2/3, 6, 0) & 2128 & 7.66294 & 7.38342 & 3.65  & 73.79 \\
			(2/3, 6, 1) & 436 & 6.07764 &  6.03979 & 0.62 & 89.07 \\
			(1, 12, 0) & 240612 & 12.3909 & 12.3009 & 0.726 & 44.11 \\
			(1, 12, 1) & 106096 & 11.5721 & 11.6461 & $-0.639$ & 45.81  \\
			(1, 12, 2) & 13856 & 9.53647 & 9.55433 & $-0.187$ & 56.18 \\
			\hline
		\end{tabular}
	\end{center}
	\vspace{0.5cm}
	\caption{
	Comparison of the statistical entropy and statistical 
	entropy at one loop for 6A orbifold.} \label{tt6a}
	\renewcommand{\arraystretch}{0.5}
\end{table}

\begin{table}[H]
	\renewcommand{\arraystretch}{0.5}
	\begin{center}
		\vspace{0.5cm}
		\begin{tabular}{|c|c|c|c|c|c|}
			\hline
			& & & & & \\
			$(Q^2,\,P^2\, Q.P)$ & $d(Q, P)$ & $S_{\rm stat}$ & $S^{(1)}_{\rm stat}$ & $\delta$ & $\delta'$ \\ 
			& & & & &   \\
			\hline
			& & & & & \\
			(1/4, 2, 0) & 20 & 2.99573 & 3.0419 & $-1.54$ & 172 \\
			(1/4, 4, 0) & 68 & 4.21951 & 4.41444 & $-4.62$ & 118 \\
			(1/4, 6, 0) & 196 & 5.27811 & 5.82516 & $-10.36$ & 88.13 \\
			(1/4, 6, 1) & 10 & 2.30259 & 2.16053 & 6.17 & 231 \\
			(1/4, 8, 0) & 504 & 6.22258 & 7.13849 & $-14.71$ & 68.82 \\
			(1/4, 8, 1) & 40 & 3.68888 &  4.06787 & $-10.27$ & 130 \\
			(3/4, 6, 0) & 2280 & 7.73193 & 7.48478 & 3.196 & 70.43 \\
			(3/4, 6, 1) & 450 & 6.10925 & 6.19909 & $-1.47$ & 83.62 \\
			(3/4, 8, 0) & 6704 & 8.81046 & 8.59292 & 2.47 & 61.47 \\
			(3/4, 8, 1) & 1728 & 7.45472 & 7.54122 & $-1.16$ & 68.57 \\
			\hline
		\end{tabular}
	\end{center}
	\vspace{0.5cm}
	\caption{
	Comparison of the statistical entropy and statistical 
	entropy at one loop for 8A orbifold.} \label{tt8a}
	\renewcommand{\arraystretch}{0.5}
\end{table}

\begin{table}[H]
	\renewcommand{\arraystretch}{0.5}
	\begin{center}
		\vspace{0.5cm}
		\begin{tabular}{|c|c|c|c|c|c|}
			\hline
			& & & & & \\
			$(Q^2,\,P^2\, Q.P)$ & $d(Q, P)$ & $S_{\rm stat}$ & $S^{(1)}_{\rm stat}$ & $\delta$ & $\delta'$ \\ 
			& & & & & \\
			\hline
			& & & & & \\
			(1/7, 2, 0) & 4 & 1.38629 & 1.5854 & $-14.3$ & 395 \\
			(1/7, 4, 0) & 10 & 2.30259 & 2.63796 & $-14.56$ & 243.8 \\
			(1/7, 6, 0)  & 24 & 3.17805 & 3.72617 & $-17.2$ & 183 \\
			(2/7, 6, 0) & 70 & 4.2485 &  4.14076 & 2.5 & 121 \\
			(2/7, 8, 0) & 156 & 5.04986 & 5.01278 & 0.73 & 103.7 \\
			(3/7, 8, 0) & 406 & 6.00635 & 5.7842 & 3.7 & 84 \\
			(5/7, 12, 0) & 11512 & 9.35115 & 9.12825 & 2.38 & 54 \\
			(5/7, 12, 1) & 4156 & 8.33231 & 8.34407 & 0.14 & 63.4  \\
			(5/7, 12, 2) & 292 & 5.67675 & 5.6847 & $-0.14$ & 93 \\
			\hline
		\end{tabular}
	\end{center}
	\vspace{0.5cm}
	\caption{
	Comparison of the statistical entropy and statistical 
	entropy at one loop for 14A/B orbifold.} \label{tt14a}
	\renewcommand{\arraystretch}{0.5}
\end{table}

\begin{table}[H]
	\renewcommand{\arraystretch}{0.5}
	\begin{center}
		\vspace{0.5cm}
		\begin{tabular}{|c|c|c|c|c|c|}
			\hline
			& & & & & \\
			$(Q^2,\,P^2\, Q.P)$ & $d(Q, P)$ & $S^{\rm stat}$ & $S^{(1)}_{\rm stat}$ & $\delta$ & $\delta'$ \\ 
			& & & & & \\
			\hline
			& & & & & \\
			(2/15, 2, 0) & 4 & 1.38629 & 1.32743 & 4.24 & 386 \\
			(2/15, 4, 0) & 8 & 2.07944 & 2.33276 & $-12.18$ & 272 \\
			(2/15, 6, 0)  & 20 & 2.99573 & 3.36918 & $-12.4$ & 193 \\
			(4/15, 6, 0) & 50 & 3.91202 & 3.79203 & 3 & 135 \\
			(4/15, 8, 0) & 102 & 4.62497 & 4.62705 & $- 0.044$ & 117 \\
			(8/15, 12, 0) & 2844 & 7.95297 & 7.76586 & 2.35 & 65.7 \\
			(8/15, 12, 1) & 898 & 6.80017 & 6.83215 & $-0.47$ & 80 \\
			(8/15, 12, 2) & 40 & 3.68888 & 3.54063 &4 & 142.8 \\
			\hline
		\end{tabular}
	\end{center}
	\vspace{0.5cm}
	\caption{Comparison of the statistical entropy and statistical 
	entropy at one loop for 15A/B orbifold.} \label{tt15a}
	\renewcommand{\arraystretch}{0.5}
\end{table}

\subsection {Exponentially suppressed corrections: the  2A orbifold }\label{2a2}

In section \ref{sec13} we have compared the contribution of the leading saddle point to the exact degeneracy 
We have shown that the constant $C_1$  present in the modular transformation 
(\ref{sp2trans}) contributes substantially to the statistical entropy at low values of charges for 
the orbifold theories. 
In this section we show that this constant is also important in the sub-leading saddles
which are exponentially suppressed compared to the leading saddle. 
For this purpose we will study 2A orbifold, a similar analysis also applies for all the 
other orbifolds $g'$ listed in table \ref{tt}. 

In \cite{Banerjee:2008ky} the first sub-leading saddle was analysed for the un-orbifolded theory 
of type II
$K3\times T^2$. 
We generalize this analysis to the case of an  orbifold of order $N$. 
Starting from the degeneracy formula (\ref{degen}) 
and using the symmetry (\ref{prop1})  and 
a change of variables we 
obtain  the following
\begin{eqnarray} \label{degen2nd}
d( Q, P) = \frac{1}{N} (-1)^{Q\cdot P +1}
\int d\tilde\rho \tilde \sigma d\tilde v 
e^{-\pi i ( \tilde\rho P^2 + \tilde\sigma Q^2 + 2 \tilde v Q. P ) }
\frac{1}{\tilde\Phi_k(\tilde\rho, \tilde\sigma, \tilde v) }
\end{eqnarray}
The contour is same as that given in (\ref{contour}). Note, however  the change
in the arguments of the exponent. 
The saddle points of the integral occur at zeros of $\tilde\Phi_k$. 
These occur at the following  hyper surfaces 
\begin{eqnarray}\label{locationzero}
n_2 ( \tilde\sigma \tilde \rho - \tilde v^2) + b \tilde v  + n_1 \tilde \sigma - 
\tilde \rho m_1  + m_2 = 0, \\ \nonumber
m_1 \in N \mathbb{Z}, \qquad  n_1 \in \mathbb{Z}, \qquad b \in 2\mathbb Z + 1, 
\qquad m_2,  n_2 \in \mathbb{Z}, \\ \nonumber
m_1n_1 + m_2 n_2 + \frac{b^2}{4} = \frac{1}{4}.
\end{eqnarray}
From the product representation of the modular form $\tilde \Phi_k$ it can be seen that
it is invariant under the transformations $\tilde\rho \rightarrow\tilde\rho+1, 
\tilde\sigma +N , \tilde v\rightarrow \tilde v+1$. 
Applying these transformations, we can bring the locations of the points
which characterize the hyper surface in (\ref{locationzero}) to 
\begin{eqnarray}\label{zeroloc}
n_1 = 0, 1\cdots (n_2-1), \\ \nonumber
m_1 = 0, N, 2N, \cdots ( n_2-1) N, \\ \nonumber
b = 1,  3,  5, \cdots 2n_2 -1.
\end{eqnarray}
Then $m_2$ is obtained by solving the last equation in (\ref{locationzero}). 

Let us now specialize to the case of $N=2$ and the second saddle $n_2=2$. 
Then using the conditions in (\ref{zeroloc}), 
the  points that characterize the hyper surface on which  
the zeros of $\tilde\Phi_k$ lie are given by 
\begin{eqnarray} \label{listzero}
(m_1, n_1, m_2, n_2,  j)_i  &=& 
\left\{  \quad (0,0,0,2,1),\,(2,0,0,2,1), \right. \\ \nonumber 
& & \quad  (0,1,0,2,1),  (2,1,-1,2,1),  \\ \nonumber
& & \quad  \,(0,0,-1,2,3),\,(2,0,-1,2,3), \\ \nonumber
& & \quad \left.  \,(0,1,-1,2,3),\,(2,1,-2,2,3).
\right\}
\end{eqnarray} 
Here the subscript $i= 1, \cdots 8$ 
labels the solution in the order written in the above equation.  For illustration, the 
location of the first zero of the second saddle $n_2 =2$  is given by 
the equation 
\begin{equation}
2 ( \tilde \rho\tilde\sigma - \tilde v^2 )  + \tilde v  =0
\end{equation}

We now need to obtain the $Sp(2, \mathbb{Z})$ transformation  that allow us to determine 
how $\tilde \Phi_k$ behaves close to these zeros. 
Let write the $Sp(2, \mathbb{Z})$ transformation more formally. 
We define  the symplectic  matrix 
\begin{eqnarray}
U_0^{-1} &=&  \left( \begin{array}{cc}
A_0  & B_0 \\
C_0 & D_0 
\end{array} \right)  \\ \nonumber
&=&  \left( \begin{array}{cccc}
1 & 0 & 0 & 0\\
1 & 0 & 0 & -1 \\
0 & - 1& 1 & 0 \\
0 & 1 & 0 & 0 
\end{array} \right) 
\end{eqnarray}
and construct 
\begin{eqnarray}
\hat \Omega_1 = ( A_0 \tilde\Omega + \tilde B_0 ) ( C_0 \tilde \Omega + D_0)^{-1} .
\end{eqnarray}
From examining this transformation  we can see that (\ref{sp2trans1})   is written
as 
\begin{equation} \label{first}
 - (i)^k C_1 \hat\Phi_k( \hat \Omega_1) =   \{ {\rm det}  ( C_0 \tilde \Omega + D_0) \}^k \tilde\Phi_k (\tilde \Omega) .
\end{equation}
Note that  the expression for $\hat v'$ in terms of $\tilde v$ does not coincide with 
any of the zeros we have in the list (\ref{listzero}). 
We need the transformation such that  $\hat \Phi_k$ is evaluated
at  these zeros. For this, we perform a further $Sp(2, \mathbb{Z})$ transformation that 
keeps $\hat \Phi_k$ invariant. Then the transformation is restricted to the following 
sub-group $G$ of $Sp(2, \mathbb{Z})$ defined by
\cite{Banerjee:2008ky}
\begin{equation}\label{four}
U_1 = \left( \begin{array}{cc} 
A_1 & B_1 \\ 
C_1 & D_1 \end{array} \right) 
\quad \rightarrow C_1 = {\bf O} \; {\rm mod\; } N, \quad 
{\rm det}\; A _1= 1\;  {\rm mod} \; N, \quad
{\rm det } \; D_1 = 1\;  {\rm mod}\;  N.
\end{equation}
Let 
\begin{equation}
\hat \Omega  = ( A_1 \hat\Omega_1 + \hat B_1 ) ( C_1\hat \Omega_1 + D_1)^{-1} ,
\end{equation}
then we define
\begin{equation}\label{second}
\Phi_k(\hat \Omega) =  \{ {\rm det}  ( C_1 \hat \Omega_1 + D_1) \}^k \Phi_k(\hat \Omega_1) .
\end{equation}
Now combining (\ref{first}) and (\ref{second})  using the group property we obtain
\begin{equation}\label{third}
-(i)^k C_1 \Phi_k(\hat \Omega) = \{ {\rm det}  ( C \tilde \Omega + D ) \}^k \tilde \Phi_k(\tilde \Omega)
\end{equation}
where
\begin{equation}
U = \left( \begin{array}{cc} 
A & B \\ 
C & D \end{array} \right) = U_1U_0^{-1}.
\end{equation} 
We can use the additional degree of freedom of performing 
a transformation within the sub-group $G$  defined in (\ref{four}) so that 
$\hat v$ as a function of $\tilde v$ coincides with the locations of the zeros given in 
(\ref{listzero}). 

We construct $U$ as follows:
\begin{enumerate}
	\item
	First take $U_1$ to be any $4\times 4$ matrix and evaluate $U_1U_0^{-1}$. 
	\item
	Evaluate the action of $U = U_1 U_0^{-1} $ on $\tilde \Omega$, we get  
	\begin{equation}
	\hat \Omega = (A \tilde\Omega + B) ( C\tilde\Omega + D)^{-1}.
	\end{equation}
	Demand that the equation for $\hat v =0 $ in terms of $\tilde v$ 
	coincides with the zeros given in  (\ref{listzero}). 
		\item
	Impose the conditions that result from $Sp(2, \mathbb{Z} )$ on $U_1$.
	\item
	Examine the resulting equations and finally impose the conditions of 
	the sub-group $G$ 
	given in (\ref{four}) on $U_1$. 
\end{enumerate}
These steps can  be implemented in Mathematica and 
the  final expression for $\hat \Omega$ written in terms of $\tilde\Omega$ 
for each the locations of the   $8$ zeros in (\ref{listzero}) are given by 
\begin{eqnarray} \label{matrices}
\hat\Omega_1&=&\left(
\begin{matrix}
\tilde\rho -\frac{\tilde v^2}{\tilde\sigma } & \frac{2 \tilde\rho  \tilde\sigma -2 \tilde v^2+
\tilde v}{\tilde\sigma } \\
\frac{2 \tilde\rho  \tilde\sigma -2 \tilde v^2+\tilde v}{\tilde \sigma } & \frac{4 \tilde\rho  \tilde
\sigma -4 (\tilde v-1) \tilde v-1}{\tilde \sigma } \\
\end{matrix}
\right) \\ \nonumber
& \\ \nonumber
\hat\Omega_2 &=& \left(
\begin{matrix}
\frac{\tilde\rho  \tilde\sigma -\tilde v^2}{8 \tilde v^2-4 \tilde v+\tilde \sigma +\tilde \rho  (4-8 \tilde \sigma )} & \frac{-2 \tilde v^2+\tilde v+2 \tilde \rho  (\tilde \sigma -1)}{8 \tilde v^2-4 \tilde v+\tilde \sigma +\tilde \rho  (4-8 \tilde \sigma )} \\
\frac{-2 \tilde v^2+\tilde v+2 \tilde \rho  (\tilde \sigma -1)}{8 \tilde v^2-4 \tilde v+\tilde \sigma +\tilde\rho  (4-8\tilde\sigma )} & \frac{-4 (\tilde v-1) v+4 \tilde \rho  \tilde \sigma -1}{8 \tilde v^2-4 \tilde v+\tilde \sigma +\tilde \rho  (4-8 \tilde \sigma )} \\
\end{matrix}\right) \\ \nonumber
& \\ \nonumber
\hat \Omega_3 &=&\left(
\begin{matrix}
\tilde \rho -\frac{\tilde v^2}{\tilde \sigma } & \frac{-2 \tilde v^2+\tilde v+2 \tilde \rho \tilde \sigma +\tilde \sigma }{\tilde \sigma } \\
\frac{-2 \tilde v^2+\tilde v+2 \tilde \rho  \tilde \sigma +\tilde \sigma }{\tilde \sigma } & \frac{-4 (\tilde v-1) \tilde v+4 \tilde \rho  \tilde \sigma +\tilde \sigma -1}{\tilde \sigma } \\
\end{matrix}
\right) \\ \nonumber
& & \\ \nonumber
\hat \Omega_4 &=& \left(
\begin{array}{cc}
\frac{\tilde v (\tilde v+2)-\tilde \rho  (\tilde \sigma -4)+1}{2 \tilde v^2-(2 \tilde \rho +1) (\tilde \sigma -2)} & \frac{\tilde v+2 \tilde \rho +1}{(2\tilde  \rho +1) (\tilde \sigma -2)-2 \tilde v^2}+1 \\
\frac{\tilde v+2 \tilde \rho +1}{(2 \tilde \rho +1) (\tilde \sigma -2)-2 \tilde v^2}+1 & \frac{2 \tilde \rho +1}{2 \tilde v^2-(2 \tilde \rho +1) (\tilde \sigma -2)}-1 \\
\end{array}
\right) \\ \nonumber
& &  \\ \nonumber
\hat \Omega_5&=&\left(
\begin{array}{cc}
\frac{-4 (\tilde v-1) \tilde v+4 \tilde \rho  \tilde \sigma -1}{\tilde\sigma } & \frac{\tilde v (3-2 \tilde v)+2 \tilde \rho  \tilde \sigma -1}{\tilde \sigma } \\
\frac{\tilde v (3-2 \tilde v)+2 \tilde \rho \tilde  \sigma -1}{\tilde \sigma } & \tilde \rho -\frac{(\tilde v-1)^2}{\tilde \sigma } \\
\end{array}
\right)
\\ \nonumber
& \\ \nonumber
\hat \Omega_6&=&\left(
\begin{array}{cc}
\frac{-4 (\tilde v-1) \tilde v+\tilde \rho  (4 \tilde \sigma -2)-1}{6 \tilde v^2-8 \tilde v+\tilde \sigma +\tilde \rho  (4-6 \tilde \sigma )+2} & \frac{\tilde v (3-2 \tilde v)+2 \tilde \rho  (\tilde \sigma -1)-1}{6 \tilde v^2-8\tilde  v+\tilde \sigma +\tilde \rho  (4-6 \tilde \sigma )+2} \\
\frac{\tilde v (3-2 \tilde v)+2 \tilde \rho  (\tilde \sigma -1)-1}{6 \tilde v^2-8 \tilde v+\tilde \sigma +\tilde \rho  (4-6 \tilde \sigma )+2} & \frac{\tilde \rho  \tilde \sigma -(\tilde v-1)^2}{6 \tilde v^2-8 \tilde  v+\tilde \sigma +\tilde \rho  (4-6 \tilde \sigma )+2} \\
\end{array}
\right)
 \\ \nonumber
 & \\ \nonumber
\hat \Omega_7&=&\left(
\begin{array}{cc}
\frac{-4 (\tilde v-1) \tilde v+4 \tilde \rho  \tilde \sigma +\tilde \sigma -1}{\tilde \sigma } & \frac{ \tilde v (3-2 \tilde v)+2 \tilde \rho \tilde \sigma +\tilde \sigma -1}{\tilde \sigma } \\
\frac{\tilde v (3-2 \tilde v)+2 \tilde \rho  \tilde \sigma +\tilde \sigma -1}{\tilde \sigma } & -\frac{(\tilde v-1)^2}{\tilde \sigma }+\tilde \rho +1 \\
\end{array}
\right)
\\ \nonumber
&  \\ \nonumber
\hat \Omega_8 &=&\left(
\begin{array}{cc}
\frac{2 \tilde \rho +1}{2 (\tilde v-2) \tilde v-2 \tilde \rho  (\tilde \sigma -2)-\tilde \sigma +4} & \frac{\tilde v-2 (\tilde \rho +1)}{2 (\tilde v-2) \tilde v-2 \tilde \rho  (\tilde \sigma -2)-\tilde \sigma +4}+1 \\
\frac{\tilde v-2 (\tilde \rho +1)}{2 (\tilde v-2) \tilde v-2 \rho  (\tilde \sigma -2)-\tilde \sigma +4}+1 & \frac{-\tilde v^2+\tilde \rho  \tilde \sigma +\tilde \sigma }{2 (\tilde v-2) \tilde v-2 \tilde \rho  (\tilde \sigma -2)-\tilde \sigma +4} \\
\end{array}
\right).
\end{eqnarray}

We can now find the behaviour of the $\tilde\Phi_k$ at these zeros. Using the 
transformation in (\ref{third})  and the factorization property of 
$\hat \Phi_k$ in (\ref{factorhatphi}), we find that 
\begin{equation}
\tilde\Phi_k(\tilde\rho,\tilde\sigma, \tilde v) |_{\rm zeros}  \rightarrow (i)^k 
[ {\rm det} (C \tilde \Omega + D)  ]^{-k} 4\pi ^2 \hat v g(\hat \rho) g(\hat \sigma) .
\end{equation}
Since we are restricting our attention to the $2A$ case, $k=6$ and 
$g(\hat\rho) = \eta^8(\rho) \eta^{8}( 2\rho)$. 
In the above equation for each of the zeros we have to substitute the 
corresponding value of $\hat \rho, \hat \sigma, \hat v$ in terms of 
$\tilde \rho, \tilde \sigma, \tilde v$
and the 
 value of $C\tilde\Omega +D$ can also be read from  (\ref{matrices}). 
The next step is to perform the contour integral over  $\tilde v$  in (\ref{degen2nd}), which will 
pick up the residue at the double pole at $\hat v =0$.  Then 
perform the saddle point integration over $\hat\rho$ and $\hat v$. 
To the leading order the saddle point is obtained by 
minimizing the exponent in (\ref{degen2nd}) given by 
\begin{equation}
E= -\pi i ( \tilde\rho P^2 + \tilde\sigma Q^2 + 2 \tilde v Q\cdot P)
\end{equation}
subject to the constraint of the location of the zero given in (\ref{locationzero}). 
The result for the saddle point is given by 
\begin{equation}
(\tilde \rho, \tilde\sigma, \tilde v) = i \left[ 2 n_2 \sqrt{ Q^2 P^2 - (Q\cdot P)^2)} \right]^{-1} ( 
Q^2, P^2, Q\cdot P )   - \frac{1}{n_2} ( n_1, -m_1, \frac{j}{2} ) .
\end{equation}
Repeating the  analysis of \cite{Banerjee:2008ky} for the and keeping track of the differences
for the case of $2A$ we see that the value at the sub-leading  saddle including the 
one loop corrections is given by 
 \begin{eqnarray}
 \Delta d( Q, P)|_{i} &=& \frac{1 }{ N C_1}  
 \frac{1}{n_2} \exp( \pi \sqrt{Q^2 P^2 - ( Q\cdot P)^2} ) \\
 \nonumber
 & & \times  \left\{ 
 [ {\rm det} ( C \tilde\Omega  +D)]^{k+2}  ( g(\hat \rho) g(\hat \sigma))^{-1}  ( 1 + O(Q^{-2}, P^{-2} ) ) 
  \right\}  \\ \nonumber
  && \left. \times (-1)^{Q\cdot P} \exp [ i \frac{\pi}{n_2} (  n_1 P^2  -m_1Q^2  - j Q\cdot P ) ]
  \right|_{\rm{saddle}_i}.
 \end{eqnarray}
 Here $n_2 = 2, k =6,$ and $g(\hat \rho) = \eta^8(\hat \rho) \eta^{8}( 2\hat \rho)$ and $N=2$. 
 The constant $C_1$ is given by $C_1 =  2^4$; \, 
 $\hat\rho, \hat\sigma$ as well as ${\rm det} ( C \tilde\Omega  +D)$ are evaluated 
 by examining the matrices given  in (\ref{matrices})  for each of the 
 $8$ saddles. This dependence on the zeros of $\tilde\Phi_k$ is  indicated by the
 subscript `$i$' \footnote{We have kept track of all signs for $k=6$. Note that we obtain 
 that sign of $j Q\cdot P$ in the second line opposite to that obtained in \cite{Banerjee:2008ky}. } . 
 The complete contribution of all saddles at $n_2 = 2$ is given by 
 \begin{equation}
 \Delta d( Q, P) = \sum_{i = 1}^8  \Delta d( Q, P)|_{i}\; .
 \end{equation}
 
 We have listed the contributions to the 2nd saddle to the degeneracies for a few low lying
 charges of dyons in the $2A$ orbifold in table \ref{2ndsaddle}.

\begin{table}[H]
	\renewcommand{\arraystretch}{0.5}
	\begin{center}
		\vspace{0.5cm}
		\begin{tabular}{|c||c|c|c|c|c|c|c|}
			\hline
			& & & & & & & \\
			$Q^2 $ & 1 & 2 & 3 & 3 & 3 & 6 & 6 \\ 
			& & & & & & & \\
			\hline
			& & & & & & & \\
			$P^2 $ & 2 & 4 & 6 & 6 & 6 & 6 & 6\\ 
			& & & & & & & \\
			\hline 
			& & & & & & & \\
			$Q\cdot P $ & 0 & 0 & 0 & 1 & 3 & 0 & 1 \\ 
			& & & & & & & \\
			\hline
				& & & & & & & \\
			$d^{(2)}$& 19.751 & 91.058 & 1679.22 & 895.668 & 0 & 83807.5 & 63637.7 \\
				& & & & & & & \\
			\hline
		\end{tabular}
	\end{center}
	\vspace{0.5cm}
	\caption{Contribution of the second saddle to the degeneracy: $\Delta d( Q, P) $ for 2A orbifold}  \label{2ndsaddle}
	\renewcommand{\arraystretch}{0.5}
\end{table}

 We mention two consistency checks on evaluating the correction  to the 
 degeneracies at the second saddle. 
 The contribution to the degeneracy from the second saddle $\Delta d( Q, P) $ has to
 be real, however the contribution from each of  $\Delta d( Q, P)|_{i}$ need not 
 be real. One of the checks is that on summing the contribution of the $8$ saddles 
 we obtain a real number. 
 The second check is that the solutions of the matrices  given in (\ref{matrices}) are not unique. 
 For example for  $5$th solution  it is also possible to choose the matrix
 consistent with all the requirements discussed earlier. 
 \begin{equation}
 \hat \Omega'_5=\left(
\begin{array}{cc}
\frac{(1-2 \tilde v)^2-4\tilde \rho  \tilde\sigma }{8 (\tilde v-1) \tilde v-(8 \tilde \rho +1) \tilde\sigma  +2} & \frac{\tilde v (3-2 \tilde v)+2 \tilde  \rho  \tilde \sigma -1}{-8 (\tilde v-1) \tilde v+8 \tilde \rho  \tilde \sigma +\tilde \sigma -2} \\
\frac{\tilde v (3-2 \tilde v)+2 \tilde \rho  \tilde \sigma -1}{-8 (\tilde v-1) \tilde v+8 \tilde \rho \tilde  \sigma +\tilde \sigma -2} & \frac{(\tilde v-2) \tilde v-\tilde \rho  (\tilde \sigma -2)+1}{8 (\tilde v-1) \tilde v-(8 \tilde \rho +1) \tilde \sigma +2}. \\
\end{array}
\right)
\end{equation}
However we have explicitly verified that the correction $\Delta d( Q, P)_5  $ 
is independent of the choice of the matrix $\hat\Omega_5$ or $\hat\Omega_5'$ for the 
charges listed in table \ref{2ndsaddle}. Finally we mention that for 
the charge $Q^2 = 3, P^2 = 6, Q\cdot P = 3$ the contribution 
to the second saddle vanishes. Such a property was also observed for the 
un-orbifolded theory in \cite{Sen:2007qy}.

In conclusion  the 
    observation that the constant $C_1$ is an important factor of all the saddle points, 
 not just the leading one at $n_1 = 1$. 
 It also occurs in the modular transformation relating $\tilde \Phi_k$ to all the other saddle points.  We have demonstrated this for the $2A$ orbifold and for the 
 saddle at $n_2 = 2$,  but it is clear from the 
 derivation that it will persist for all the other sub-leading saddles as well 
 the other orbifolds. 
 Finally we would also like to mention that the contribution of all  the sub-leading saddles
 for the case of the un-orbifolded  theory was evaluated in \cite{Murthy:2009dq}. 
 This was done by a characterization of all the solutions of the zeros of $\Phi_{10}$ for  all 
 the sub-leading saddles by  determining the $Sp(2, \mathbb{Z})$ transformation which maps the 
 zeros to the zero at $v=0$. It will be interesting to repeat this analysis for the $\mathbb{Z}_N$ orbifolds
 considered in this paper. Performing this would lead to a Farey tale like  expansion of the
dyon partition function for these theories. This would be important to understand the sub-leading 
contributions from semi-classical geometry.

\subsection{Implications of the constant $C_1$} \label{sec2loc}

We briefly discuss the implication of the presence of the non-trivial constant  $C_1$ 
in the one loop saddle point approximation to the entropy  $S_{\rm stat}^{(1)}$. 
In the large charge limit, the $1/4$ BPS dyon can be described by 
an extremal black hole with the same charge $(Q, P)$. 
We can evaluate the Wald's generalization of the Bekenstein-Hawking entropy to this black hole. 
This is done by considering the four derivative correction to the effective action 
of the ${\cal N }=4$ string theory given by the Gauss-Bonnet term \footnote{ See
\cite{Sen:2007qy} for a review. }

\be
{\cal L}=\phi(a, S)\left(R_{\mu\nu\rho\sigma}R^{\mu\nu\rho\sigma}-4R_{\mu\nu}R^{\mu\nu}+R^2\right),
\ee
where $a, S$ is the axion and dilaton respectively in the heterotic description of the theory. 
Using the analysis of \cite{Gregori:1997hi,David:2006ud} we can compute the function $\phi$. 
It is given by 
\begin{eqnarray} \label{gaussb}
\phi(a, S)  &=& -\frac{1}{64\pi^2} \left[ (k+2) \ln S + \ln g( a + i S)  + \ln g(  - a + i S)  \right]  \\ \nonumber
& & + {\rm constant}
\end{eqnarray}
Here $g$ is the  same function  that occurs in the statistical entropy function given in (\ref{statent}) and listed in 
table \ref{t6} for all the orbifolds. 
However the constant present in (\ref{gaussb}) cannot be fixed. 
This is because the  a constant coefficient in the Gauss-Bonnet term is a total derivative and therefore 
does not affect the equations of motion. 
The evaluation of the Gauss-Bonnet term is done through a string amplitude calculation which can determine 
the effective action only up to on shell terms.
 A constant in Gauss-Bonnet  cannot be determined using this procedure. 
 Now going evaluating the Wald entropy of the extremal dyonic black hole, one obtains the same 
 minimizing problem as encountered in (\ref{statent}). That is 
 \begin{eqnarray}
 S(a, S)  &=& \frac{\pi}{2 S} | Q + ( a+ i S) P|^2  -\ln g( a + i S) - \ln g( - a + i S) - (k+2) \ln ( 2 S)  \nonumber \\
 & &   + {\rm constant}.
 \end{eqnarray}
 The Wald entropy is given by 
 \begin{eqnarray}
 S_{\rm Wald} =  S(a, S)|_{\rm minimum}\; .
 \end{eqnarray}
 Therefore the undetermined constant in the Gauss-Bonnet term turns out to be the $O(Q^0, P^0)$ term
 in the Wald entropy. 
 Thus it is cleat that a string amplitude calculation will not be able to fix this constant  coefficient in the Gauss-Bonnet
 term.  
 
 Recently beginning with the works of \cite{Dabholkar:2010uh,Dabholkar:2011ec} 
 the method of localization in $AdS_2\times S^2$ has 
 proposed to evaluate the entropy of these black holes exactly from a geometric description.
 All of the works so far address only the un-orbifolded theory of compactification on $K3\times T^2$. 
 From the analysis in this paper it is clear that the constant $C_1$ substantially contributes to the 
 entropy at low charges for all the orbifolds. We see that it is a $O(Q^0, P^0)$ term. The method of 
 localization is exact and is hoped that partition functions agree  with the microscopic description at finite order
 in charges. 
 Therefore an important test for the method of 
 localization is to reproduce this constant $C_1$.

\section{ Fourier-Jacobi coefficients of $1/\tilde\Phi_k$  } \label{sec3}

Consider the following Fourier expansion if the partition function of 
dyons of the un-orbifolded theory, ie. type II compactified on $K3\times T^2$
\footnote{We have changed variables 
$\rho \rightarrow \tau$  and $v\rightarrow z$ to confirm with the 
standard notation for Jacobi forms. }
\begin{equation}\label{fjdecomp}
\frac{1}{\Phi_{10}  ( q,  p, y ) }  =   \sum_{m = -1}^\infty \psi_m (\tau, z )  p^m, 
\qquad q = e^{2\pi i \tau}, p = e^{2 \pi i \sigma}, y = e^{2\pi i z} .
\end{equation}
The Fourier-Jacobi coefficients $\psi_m$  after multiplying by 
$\eta^{24} (\tau) $ are meromorphic Jacobi forms of
weight $2 $ and  index $m$.  Note that the 
Fourier-Jacobi coefficients count degeneracies  for a given magnetic charge of the 
dyon. 
In \cite{Dabholkar:2012nd} it was shown that 
 the Fourier-Jacobi coefficients $\psi_m$ for $m>0$  admits
the following  unique decomposition 
\begin{equation} \label{splitting}
\psi_m(\rho, z)  = \psi_m^{\rm P} (\tau, z)  + \psi_m^{\rm F} (\tau, z).
\end{equation}
in which the polar part $\psi_m^P(\tau, z)$ has the same pole structure in  the $z$-plane
as 
$\psi_m(\tau, z)$ and $\psi_m^{\rm F} (\tau, z)$ has no poles.  The polar part 
is given by  the following Appell-Lerch sum 
\begin{eqnarray}
\psi_m^{\rm P} ( \tau, z) = \frac{p_{24}(m+1)}{\eta^{24}(\tau) } {\cal A}_{2, m }(\tau, z) , 
\\ \nonumber
{\cal A}_{2, m }(\tau, z) = \sum_{s\in \mathbb{Z}} 
\frac{q^{ms^2 +s}  y^{2ms +1}}{ (1 - q^s y )^2}  .
\end{eqnarray}
The   Appell-Lerch sum exhibits wall-crossing and therefore 
$\psi_m^{\rm F}$  capture the degeneracies of multi-centered black holes. 
The Fourier coefficients of $\psi_m^{\rm F} (\tau, z )$ are independent of the 
choice of contour in the $(z,\tau)$ space and counts the degeneracies of 
immortal black holes. 
Further more $\psi_m^{\rm F} (\tau, z )$ is a mock Jacobi form. 

In this section we generalize these observations for the  partition function of 
dyons in type II compactifications on $K3\times T^2/Z_{\mathbb{N}}$, where 
the orbifold is performed by the action of $g'$ given in table \ref{tt}. 
This observation is made for the  2 lowest values of  magnetic charges. 
In this case $\psi_m$ after multiplying by an appropriate $\Gamma_0(N)$
form are meromorphic Jacobi forms under $\Gamma_0(N)$. 
We will begin with the case of $2A$ and then move to  all the other
orbifolds.

\subsection{The 2A orbifold}

The  relevant  Siegel modular form for the $2A$ orbifold  is $\tilde \Phi_6$. 
In the product form given in (\ref{siegform}), the input required for its construction is 
given by the twisted elliptic genus of the $2A$  orbifold which is given by 
\begin{eqnarray}\label{2atwist}
F^{(0, 0)} &=& 4 A(\tau, z) , \\ \nonumber
F^{(0, 1)} &=& \frac{4}{3} A(\tau, z) - \frac{2}{3} B(\tau, z) {\cal E}_2(\tau) , \\ \nonumber
F^{(1, 0)} &=& \frac{4}{3} A(\tau, z) + \frac{1}{3} B(\tau, z) {\cal E}_2(\frac{\tau}{2} ) , 
\\ \nonumber
F^{(1, 1)} &=& \frac{4}{3}A(\tau, z) + \frac{1}{3} B(\tau, z) {\cal E}_2( \frac{\tau +1}{2} ) .
\end{eqnarray}
Here $A$ and $B$ are defined in (\ref{defab}) and ${\cal E}_N$ is defined as 
\begin{equation}
{\cal E}_N(\tau) = N E_2( N\tau) - E_2( \tau) .
\end{equation}
Using the expansions of the twisted elliptic genus $c^{(r, s)}$ as defined in 
(\ref{expelipg}) we can construct  $\tilde \Phi_6$. 
It is easy to see that for the $2A$ orbifold, the expansions  for the inverse of 
$\tilde\Phi_6$ is given by 
\begin{equation}\label{exphi6}
\frac{1}{\tilde\Phi_6( q, p, y) } = \sum_{m = -1}^\infty \psi_{m}( \tau, z)  p^{\frac{m}{2} }.
\end{equation}
Note that the expansion in terms of the magnetic variable $y$ can now take 
half integral values.  Furthermore  from (\ref{degen}) we see that $m$ is related to the
magnetic charge by $P^2 = 2m$. 
From the explicit construction of $\Phi_6$ and the expansion in terms of $y$, we can 
read out the following
\begin{eqnarray}\label{wtjacobi}
[ \eta( \tau) \eta( 2\tau) ]^8  \psi_{-1} &=& -  \frac{1}{B(2\tau, z) }, \\ \nonumber
[ \eta( \tau) \eta( 2\tau) ]^8  \psi_{0} &=& -  2 \frac{F^{(1, 0)} ( 2\tau, z) }{ B(2\tau, z) }, 
\\ \nonumber
[ \eta( \tau) \eta( 2\tau) ]^8  \psi_{1} &=&  - \frac{1}{B( 2\tau, z) }  \\ \nonumber
& & \times \left[
F^{(0,0)}(\tau,z)+F^{(0,1)}(\tau+1/2,z)+2[F^{(1,0)}(2\tau,z)]^2+F^{(1,0)}(4\tau,z^2) \right].
\end{eqnarray}
We define
\begin{equation}
g^{(8)} ( \tau ) = [ \eta( \tau) \eta( 2\tau) ]^8 (\tau).
\end{equation}
The next step is  to follow the procedure of \cite{Dabholkar:2012nd} and write down an Appell-Lerch sum
whose poles and residues coincide with the weight $2$ Jacobi forms transforming 
under $\Gamma_0(2)$ on the left hand side of (\ref{wtjacobi}). 
We do not consider $m = -1$, since the Appell-Lerch sum diverges. 
This is easy to see,  
because the  meromorphic Jacobi form $1/B(2\tau, z)$ is identical  to the function for the 
un-orbifolded theory with the replacement of $\tau \rightarrow 2\tau$.

{\bf $P^2 = 0, m = 0$}

Let us examine the case $m=0$. Using the expression for the twisted elliptic 
genus in (\ref{2atwist}) we see that it reduces to 
\begin{equation} \label{p1}
- g^{(8)} (\tau) \psi_{0} =  \frac{8}{3} \frac{A(2\tau, z)}{B(2\tau, z) }
+ \frac{2}{3} {\cal E}_2( \tau) .
\end{equation}
Now we can re-write 
\begin{equation}\label{p2}
4\frac{A(2\tau, z)}{B(2\tau, z) }  =  -12 \sum_{n\in \mathbb{Z}} \frac{ q^{2n} y}{ ( 1- q^{2n} y ) ^2 }
     - E_2 ( 2\tau) .
\end{equation}
where $E_2$ is the non-holomorphic Eisenstein series of weight $2$. 
This identity was used in \cite{Dabholkar:2012nd} \footnote{Let us remark about the notation
of \cite{Dabholkar:2012nd} in comparison with ours. $4 A(\tau, z)_{\rm ours} =  B(\tau, z)_{\rm theirs}, \;
 B(\tau, z)_{\rm ours} = - A(\tau, z)_{\rm theirs}$ }. 
Combining (\ref{p1}) and (\ref{p2}) we see that 
that the polar  and the finite part is given by 
\begin{eqnarray} \label{2afinite}
g^{(8)} (\tau)\psi_{0}^{\rm P} =  8 \sum_{n\in \mathbb{Z} }\frac{ q^{2n}y }{ ( 1- q^{2s} y ) ^2 }, \\ \nonumber
g^{(8)} (\tau)  \psi_{0}^{\rm F}  = \frac{2}{3} \left(  E_2 ( 2\tau)  -  {\cal E}_2( \tau)  
\right) .
\end{eqnarray}
Therefore we have decomposed the meromorphic Jacobi form that occurs in the expansion 
(\ref{exphi6}) at $m=0$ to a polar part and a finite term.  The finite term contains the mock modular 
form $E_2 (2\tau) $ as
well as ${\cal E}_2(\tau)$  both of which transforms under $\Gamma_0(2)$ with weight $2$.  
In appendix \ref{mock}
we show that ${E}_2(2\tau)$ is a Mock modular form of weight 2 in $\Gamma_0(2)$.

{\bf $P^2 = 2, m = 1$}

The first step  in the analysis of the  meromorphic Jacobi from that occurs at $m=1$ 
is to use the equations in the expression of the twisted elliptic genus  (\ref{2atwist})  to obtain
\begin{eqnarray} \nonumber 
-  g^{(8)}(\tau) \psi_{1} &=& 
\frac{44}{9}\frac{A^2(2\tau,z)}{B(2\tau,z)}+\frac{8}{3}\frac{A(\tau,z)}{B(2\tau,z)}+\frac{16}{9}A(2\tau,z){\cal E}_2(\tau)+\frac{2}{9}B(2\tau,z){\cal E}_2^2(\tau)\\ \nn
&&+\frac{1}{4}E_4(2\tau)B(2\tau,z)+\frac{4}{3}\frac{\theta_2^2(2\tau,z)}{\theta_2^2(2\tau)}{\cal E}_2(2\tau)-\frac{2}{3}\frac{\theta_3^2(2\tau,z)}{\theta_3^2(2\tau)}{\cal E}_2(\tau+1/2) .\\  \label{p3}
\end{eqnarray}
Here we have also used the identities 
\begin{eqnarray}\label{rel}
&& \frac{3}{4}E_4(\tau)B(\tau,z)^2=4\left( A(2\tau, 2z)+A(\frac{\tau}{2},z)+A(\frac{\tau+1}{2},z)-A^2(\tau,z)\right)\\ \nn
&& \frac{B(2\tau,z^2)}{B(\tau,z)} = 4\frac{\theta_2^2(\tau,z)}{\theta_2^2(\tau)},\qquad \frac{B(\frac{\tau}{2}, 2z)}{B(\tau,z)} = \frac{\theta_4^2(\tau,z)}{\theta_4^2(\tau)}, \qquad \frac{B( \frac{\tau+1}{2}, 2z)}{B(\tau,z)} = \frac{\theta_3^2(\tau,z)}{\theta_3^2(\tau)}.\\ \nn
\end{eqnarray}
It is more convenient to convert the ratios of theta functions in (\ref{p3}) to Jacobi forms $A$ and $B$. 
For this we use the identities
\begin{eqnarray}
4\frac{\theta_2(\tau,z)^2}{\theta_2^2(\tau)}&=&\frac{4}{3}A(\tau,z)-\frac{2}{3}B(\tau,z){\cal E}_2(\tau),\\ \nn
4\frac{\theta_4(\tau,z)^2}{\theta_4^2(\tau)}&=&\frac{4}{3}A(\tau,z)+\frac{1}{3}B(\tau,z){\cal E}_2(\tau/2),\\ \nn
4\frac{\theta_3(\tau,z)^2}{\theta_3^2(\tau)}&=&\frac{4}{3}A(\tau,z)+\frac{1}{3}B(\tau,z){\cal E}_2((\tau+1)/2).
\end{eqnarray}
Thus  $\ref{p3} $ becomes
\begin{eqnarray} \nonumber
-  g^{(8)}(\tau)  \psi_{1} &=& 
\frac{44}{9}\frac{A^2(2\tau,z)}{B(2\tau,z)}+\frac{8}{3}\frac{A(\tau,z)}{B(2\tau,z)}+\frac{16}{9}A(2\tau,z){\cal E}_2(\tau)+\frac{2}{9}B(2\tau,z){\cal E}_2^2(\tau)\\ \nn
&&+\frac{1}{4}E_4(2\tau)B(2\tau,z)+\frac{1}{3}\left[\frac{4}{3}A(2\tau,z)-\frac{2}{3}B(2\tau,z){\cal E}_2(2\tau) \right]
{\cal E}_2(2\tau)\\ \nn
&&-\frac{1}{6}\left[ \frac{4}{3}A(2\tau,z)+\frac{1}{3}B(2\tau,z){\cal E}_2(\tau+1/2)\right]{\cal E}_2(\tau+1/2) .\\ \label{p4}
\end{eqnarray}

It is clear from this expression that  polar terms  arise from the meromorphic 
Jacobi forms ${A^2(2\tau,z)}/{B(2\tau,z)}$ and 
${A(\tau,z)}/{B(2\tau,z)}$  of weight $2$ transforming under $\Gamma_0(2)$ with index $1/2$. 
We can use the following identity derived in \cite{Dabholkar:2012nd} to re-write ${A^2(2\tau,z)}/{B(2\tau,z)}$ 
into an Appell-Lerch sum and a mock modular form
\begin{equation}\label{appel2}
-16 \frac{A^2 ( 2\tau, z) }{B(2\tau, z)} = 144 \sum_{n \in \mathbb{Z}} \frac{ q^{2n^2 +2n} y^{2n +1}}{ (1- q^{2n} y)^2 } -
E_4(2\tau) B(2\tau, z)  - 288 {\cal H} (2\tau, z) .
\end{equation}
Here ${\cal H}$ is  the simplest  Jacobi mock modular form related to the generating function of 
Hurwitz-Kronecker class numbers
\begin{eqnarray}
{\cal H}(\tau, z)  = \sum_{n =0}^\infty   H( 4n - j^2) q^n y^l .
\end{eqnarray}
The coefficients $H(n)$ are defined by 
\begin{eqnarray}
H( n)  &=& 0 \qquad \hbox{for}  \; n <0, \\ 
 \sum_{ n\in\mathbb{Z} }  H(n) q^n &=& 
-\frac{1}{12} +\frac{1}{3} q^3  + \frac{1}{2} q^4 + q^7 + q^8  + q^{11}  + \cdots
\end{eqnarray}
What remains is to figure out how to write the meromorphic Jacobi form ${A(\tau,z)}/{B(2\tau,z)}$ as a polar part 
and a finite term. 
Note that the location of the poles of order $2$ lie precisely at the same point as the form 
${A^2(2\tau,z)}/{B(2\tau,z)}$. Further more some bit of analysis shows that the residue at the double pole and the 
simple pole of  $ 3\frac{A(\tau,z)}{B(2\tau,z)}$ is precisely  equal to  the corresponding residues  of 
${A^2(2\tau,z)}/{B(2\tau,z)}$. 
A bit more study show that we can derive the following identity satisfied by the meromorphic Jacobi forms
\begin{eqnarray}\label{basicident}
\frac{A(\tau,z)}{B(2\tau,z)} &=& \frac{1}{3}\frac{A^2(2\tau,z)}{B(2\tau,z)}+\frac{1}{12}A(2\tau,z){\cal E}_2(\tau)+\frac{1}{80}B(2\tau,z)E_4(\tau)\\ \nn
&&-\frac{1}{80}B(2\tau,z)E_4(2\tau).
\end{eqnarray}
Further more we have the identities
\begin{eqnarray}\label{gamma02ident}
&&-2 {\cal E}_2(\tau) + {\cal E}_2(\frac{\tau}{2} ) + {\cal E} ( \frac{\tau+1}{2} ) = 0 , \\ \nonumber
&&E_4(2\tau)=\frac{-1}{4}E_4(\tau)+\frac{5}{4}{\cal E}_2^2(\tau), \\ \nonumber
&&{\cal E}_2(2\tau){\cal E}_2(\tau+1/2)=\frac{-3}{8}E_4(\tau)+\frac{11}{8}{\cal E}_2^2(\tau), \\ \nonumber
&&4{\cal E}_2^2(2\tau)+{\cal E}_2^2(\tau+1/2)=\frac{13}{2} {\cal E}_2^2(\tau)-\frac{3}{2}  E_4(\tau) .
\end{eqnarray}
Now  substituting (\ref{basicident}), (\ref{gamma02ident}) in (\ref{p4}) we obtain
\begin{eqnarray}\nn
- g^{(8)} (\tau) \psi_{1} &=& \frac{52}{9}\frac{A^2(2\tau,z)}{B(2\tau,z)}+\frac{20}{9}A(2\tau,z){\cal E}_2(\tau)+B(2\tau,z)\left(\frac{1}{16}E_4(\tau)+\frac{19}{144}{\cal E}_2^2(\tau)\right)
\end{eqnarray}
Finally using the expansion  in (\ref{appel2}) and the identities in (\ref{gamma02ident})  we obtain 
\begin{eqnarray} \label{decomp}
& &g^{(8)} (\tau) \psi_{1}^{\rm P}   = 52 \sum_{n \in \mathbb{Z}} \frac{ q^{2n^2 +2n } y^{2n +1}}{ (1- q^{2n} y)^2 } 
\\ \nonumber
&& g^{(8)} (\tau) \psi_{1}^{\rm F} = - \frac{20}{9 } A( 2\tau, z) {\cal E}_2(2\tau)
-  B(2\tau, z)  \left( \frac{13}{36} E_4(2\tau) + \frac{1}{16} E_4(\tau) + \frac{19}{144} {\cal E}_2^2 (\tau) \right) 
\\ \nonumber
& & \qquad\qquad\qquad\qquad \qquad - 104 {\cal H} ( 2\tau, z) 
\end{eqnarray}
From the final expression for the finite part  $\psi_1^F$ we see that the relevant mock modular forms
is the one constructed from the Hurwitz-Kronecker class numbers. It is the ``optimal'' choice for the mock modular
form.  This is because the Fourier coefficients of $q^n y^l$ of the expansion of ${\cal H}( q, z)$ grow 
polynomially in $4n - l^2$.  
In appendix \ref{mock}
we show that ${\cal H}(N\tau,z)$ is a Mock Jacobi form of index $1/N$ in $\Gamma_0(N)$.

One important observation from our analysis at $m=1$ is that   that there existed 
a different meromorphic Jacobi form $A(\tau, z) /B(2\tau, z)$ at the intermediate steps of our analysis. 
However the identity (\ref{basicident}) related it to the form $A^2(2\tau, z) /B(2\tau, z)$.
We could then use the identities obtained by \cite{Dabholkar:2012nd} to  obtain the polar and finite parts of the 
Fourier-Jacobi coefficient of $1/\tilde\Phi_6(\rho, \sigma, v)$ at $m=1$. 
Though we  demonstrated  this feature only till $m=1$,   our preliminary analysis  indicates that 
this persists in the expansion 
at higher magnetic charges and the identities obtained by \cite{Dabholkar:2012nd} are sufficient to obtain 
the mock modular forms that determine the polar and the finite parts.  

\subsection*{Fourier coefficients of $\psi_1^{\rm F}$ and single centered dyons}

The decomposition  of $\psi_1$ given in (\ref{decomp})  has physical implications. 
Note that there is no ambiguity in  the Fourier expansion of $\psi_1^{\rm F} $, while 
the Fourier expansion of $\psi_1^{\rm P} $ depends on the domain in the space  $(q, y)$ where the 
expansion is performed.  Let us elaborate on this further. 
We have defined the degeneracy using the contour defined in (\ref{contour}). 
The ensures that we are in the region ${\cal R}$, right of the line that joins 
$0$ and $i\infty$  in the axion-dilaton moduli space
The region ${\cal R}$ was found in \cite{Sen:2007vb} and is  shown in figure 1. of \cite{Sen:2010mz}.
For completeness we have  included  the figure 1 of \cite{Sen:2010mz}  below. 
\begin{figure}[hbtp]
	\centering
	\includegraphics[scale=0.5]{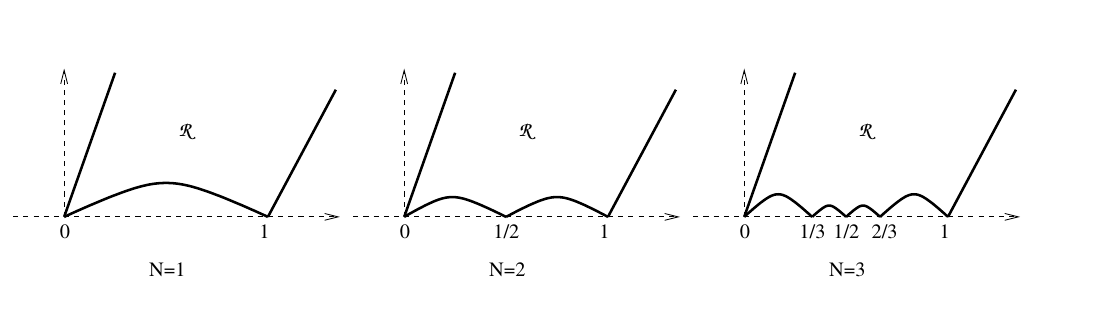}
	\caption{Figure reproduced from \cite{Sen:2010mz} showing 
	  chamber ${\cal R}$ in the upper half axion-dilaton plane, bounded by the walls of marginal stability, for the un-orbifolded theory, the $2A$ and the $3A$ orbifolds.}\label{cont}
	\end{figure}
Using the definitions in (\ref{fjdecomp}) we see that we can first perform the 
expansions in $p$ and therefore obtain the partition function $\psi_1$ for fixed magnetic 
charge $P^2 = 2$. 
Now from the contour defined in (\ref{contour}) we see that 
\begin{equation}
|q| << 1, \quad \frac{1}{|y|} <<1, \quad  |q^n y| <<1, \qquad |\frac{q^n}{y} | <<1
\end{equation}
where $n>0$.  This implies that we expand the polar part as 
\begin{eqnarray}
g^{(8)}(\tau) \psi_1^{\rm P}  = 52 \left( \cdots \frac{ q^4 y^{-3}}{ ( 1 - \frac{q^2}{y} )^2}
+ \frac{1 }{ y  ( 1- \frac{1}{y} )^2} 
+ \frac{ q^4 y^3 }{ ( 1- q^2 y) } + \cdots \right) 
\end{eqnarray}
We can then further expand the denominator in each of the terms. 
It is clear from this expansion, which is determined by the 
contour in (\ref{fjdecomp})  that the polar part contributes only to  the degeneracies 
with $Q\cdot P \leq -1$ and $Q\cdot P \geq 3$. 
Therefore the degeneracies for $Q\cdot P =0, Q\cdot P = 1, Q\cdot P = 2$ with $P^2 = 2$ and 
$Q^2 \geq 1/2$ 
are entirely determined by $\psi_1^{\rm F} $ for the contour choice in (\ref{contour}). 

Let us now examine the kinematic constraints on charges found by \cite{Sen:2010mz} 
which ensures that the 
attractor moduli for single centered black holes line in the region ${\cal R}$. 
For the $2A$ orbifold the conditions are  given by \footnote{See  inequalities in equation 3.9 along 
along with the  discussion around equation 3.5 of \cite{Sen:2010mz}.}
\begin{eqnarray} \label{range}
Q^2>0;\quad P^2>0; \quad Q^2 P^2>(Q.P)^2;\quad 2Q^2+P^2-3(Q\cdot P )^2\ge 0;\\ \nn
2Q^2 \ge Q\cdot P; \quad P^2 \ge Q\cdot P; \quad Q\cdot P\ge 0.
\end{eqnarray}
For $P^2 =2$, the  condition $Q\cdot P \leq P^2$
 tell us that $0\leq Q\cdot P \leq 2$. 
This fact combined with the discussion in the previous  paragraph 
allows us to conclude that is the Fourier coefficients of $\psi_1^{\rm F}$ which 
determines the degeneracies of the single centered black hole and  $\psi_1^{\rm P}$ does not 
contribute to these degeneracies. 
Note however that $\psi_1^{\rm F}$ also contains degeneracies for example
 $Q^2=1/2, Q\cdot P  = 2, P^2 = 2$. 
This however does not satisfy the constraint $( Q\cdot P)^2 < Q^2 P^2$ or the constraint $2Q^2 \geq Q\cdot P$. 
{\it Thus we conclude:
the degeneracies of the single centered dyons are contained in $\psi_1^F$, but not all 
Fourier coefficients of $\psi_1^F$ represent single centered dyons. }

A similar conclusion was also reached  in \cite{Dabholkar:2012nd} for the un-orbifolded theory by considering 
the attractor contour.  In that work an aprioi definition of  $\psi_m^{\rm F}$ was  given in terms
of a contour originally introduced by \cite{Cheng:2007ch}. 
The attractor contour  extracts  the 
degeneracies of single centered black holes. Here we have defined $\psi_1^{\rm F}$ in terms of the 
splitting equation (\ref{splitting}) and (\ref{decomp}). 
The crucial meromorphic form that occurs at this level which is given by $A^2( 2\tau, z) /B(2\tau, z)$
is the same as that found 
in \cite{Dabholkar:2012nd} but with the argument  $\tau\rightarrow 2\tau$. 
Thus one can indeed carry provide an a priori definition of $\psi_m^{\rm F}$ by following the steps of 
\cite{Dabholkar:2012nd}.  We leave this for future work. As we will subsequently 
see,  for the discussions in the rest of the paper, 
the above analysis  indicates that $\psi_1^F$ contains the degeneracy of the 
single centered black holes is sufficient.

\subsection{ Other orbifolds $g'$  of $K3\times T^2$ }

In this section we study the Fourier-Jacobi coefficients of $1/\tilde\Phi_k$ of all the other orbifolds listed
in table \ref{tt}. Our objective is to demonstrate that the mock modular forms which determine the 
finite part of the Fourier-Jacobi coefficients till $m=1$ is that same as that for the un-orbifolded theory. 
The Fourier-Jacobi coefficients of $1/\tilde\Phi_k$  are defined as
\begin{equation}
\frac{1}{\tilde\Phi_k(q, p, y)} = \sum_{m =-1}^\infty \phi_m (\tau, z) p^{\frac{m}{N} }
\end{equation}
We restrict our analysis to $m = -1, 0, 1$. 
The first step is to use the product form given in (\ref{siegform}) to obtain $\psi_m^{(g')}$. 
We divide our analysis for  $g'$ whose order  $N$ is odd and $g'$ whose order is even. 

\vspace{.5cm} \noindent
{\bf $g'$ with order  $N$ odd}
\vspace{.5cm}

This case includes  the orbifolds $pA$ with $p = 3, 5, 7, 11$ and $15A$ and $15B$. 
From the product form given in (\ref{siegform})  
we obtain 
\begin{eqnarray} \label{magexp1}
-\psi_{-1} &=& \frac{1}{B(N\tau,z) g^{(k+2)} (\tau) } \\ \nonumber
-\psi_{0} &=& \frac{ NF^{(1,0)}(N\tau,z)}{ B(N\tau,z) g^{(k+2)}(\tau) } , \\ \nonumber
-\psi_{1}  &=& \frac{1}{B(N\tau,z) g^{(k+2)}(\tau)  }\times \\ \nonumber
&&\left[(\frac{N^2}{2}F^{(1,0)}(N\tau,z))^2+\frac{N}{2}\left(F^{(1,0)}(2N\tau, 2 z)+F^{(2,0)}(\frac{N\tau}{2},z)+F^{(2,0)}(\frac{N(\tau+1)}{2},z)\right)\right].
\end{eqnarray}
Here the $\Gamma_0(N)$ form $g^{(k+2)} (\tau) $  of weight $k+2$ for each case can be read out from 
the table \ref{t6}.  The weights can be read out from table \ref{t3}. 
For example for $pA$ it is given by $g^{(k+2)} (\tau) = \eta^{k+2}(\tau)\eta^{k+2}(p\tau)$. 
Now we can use the from of the twisted elliptic genus in (\ref{explielip}) to separate out the meromorphic 
Jacobi form. 
For $m=0$ we obtain 
\begin{equation}
- g^{(k+2)} \psi_{0}
 =  \alpha_{g'}^{(1,0)} \frac{A(N\tau, z) }{ B(N\tau, z) } + N \beta_{g'}^{(1,0)}( N\tau) .
\end{equation}
Recall the constants $\alpha_{g'}^{(r,s)}$  and the $\Gamma_0(N)$ form $\beta_{g'}^{(r,s)}(\tau)$ are 
read out from the explicit computation of the twisted elliptic genus for all the orbifolds in table \ref{tt} from 
\cite{Chattopadhyaya:2017ews}. 
Using the identity in (\ref{p2}) we see that the polar and the finite parts of $\psi_0$ are given by 
\begin{eqnarray} \label{polfint}
g^{(k+2)} \psi_0^P &=& 3 N \alpha^{(1, 0)} \sum_{n \in \mathbb{Z}} \frac{q^{Nn} y}{ ( 1- q^{Nn } y)^2}, \\ \nonumber
g^{(k+2)}\psi_0^F &=& \frac{ N \alpha^{(1, 0)}}{4} E_2( N\tau) - N \beta^{(1, 0)}( N\tau) .
\end{eqnarray}
For $m=1$ we use the identities in (\ref{rel})  to isolate the meromorphic Jacobi form of weight $2$ and index
$1/N$ transforming under $\Gamma_0(N)$. 
This results in 
\begin{eqnarray}\label{p8}
& & - g^{(k+2)} \psi_1 = \frac{N}{2}\alpha^{(1,0)}_{g'} (\frac{N}{2}\alpha^{(1,0)}_{g'}+1)\frac{A^2(N\tau,z)}{B(N\tau,z)}
+ \frac{N}{2}\alpha^{(1,0)}\frac{3}{16}E_4(N\tau)B(N\tau,z)+
\\ \nn
&&\qquad\qquad +N^2\alpha^{(1,0)}_{g'}
A(N\tau,z)\beta^{(1,0)}(N\tau)+\frac{N^2}{2}\beta^{(1,0)}_{g'} (N\tau)^2 B(N\tau,z)\\ \nn
&&+\frac{N}{2}\left[4\frac{\theta_2(N\tau,z)^2}{\theta_2^2(N\tau)}\beta^{(1,0)}_{g'} (2N\tau)+\frac{\theta_3(N\tau,z)^2}{\theta_3^2(N\tau)}\beta^{(1,0)}_{g'} (\frac{N(\tau+1)}{2} )+\frac{\theta_4(N\tau,z)^2}{\theta_4^2(N\tau)}\beta^{(1,0)}_{g'}
(\frac{N\tau}{2})\right)].
\end{eqnarray}
From the above expression, we see that the only meromorphic Jacobi form is 
is the term with $A^2(N\tau, z) /B(N\tau, z) $.  We can use the identity in (\ref{appel2}), to write this form 
as a polar part and a finite term. 
Also we can see that the rest of the terms in (\ref{p8}) can in principle be written in terms of Jacobi forms
transforming under $\Gamma_0(N)$.  The polar term is given by 
\begin{eqnarray} 
g^{(k +2)} \psi_1 = \frac{9N}{2}\alpha^{(1,0)}_{g'}(\frac{N}{2}\alpha^{(1,0)}_{g'}+1) 
 \sum_{n \in \mathbb{Z}} \frac{ q^{Nn^2} y^{Nn +1}}{ 1- q^{Nn} y }.
 \end{eqnarray}
 It is not illuminating to write the finite term explicitly, but it can be written down if needed. 
 We note that the mock modular form that appears
 in the finite term is  essentially the generating function of Hurwitz-Kronecker class numbers ${\cal H}(N\tau, z) $.

\vspace{.5cm} \noindent
{\bf $g'$ with order  $N$ even}
\vspace{.5cm}

Here we deal with the orbifolds belonging to  class $4B, 6A, 8A, 14A, 14B$. 
Again examining the product representation (\ref{siegform}) we obtain
\begin{eqnarray}
-\psi_{-1}(q,z)&=&\frac{1}{B(N\tau,z) g^{(k+2)} (\tau) }, \\ \nn
-\psi_0(q,z)&=&\frac{NF^{(1,0)}(N\tau,z)}{ B(N\tau,z)  g^{(k+2)} (\tau) }. \\ \nn
\end{eqnarray}
Comparing the 
 $m=0$ expression with that of (\ref{magexp1}) we see that this is same as that when $N$ is odd.
 Therefore 
 the analysis proceeds identically   and we obtain 
(\ref{polfint}) for the polar and the finite part. 
For $m=1$ all the orbifolds with $N$ even yield different expressions
\begin{eqnarray} \nonumber
{\; 4B:\;}- g^{(k+2)} (\tau)\psi_1&=& (8F^{(1,0)}(4\tau,z))^2 \\ \nonumber
&& +2\left[F^{(1,0)}(8\tau,  2z)+F^{(2,0)}(2\tau,z)+F^{(2,1)}\left(2\tau + \frac{1}{2} ,z\right)\right],\\ \nn
{6A:\;} -g^{(k+2)} (\tau) \psi_1&=& (18F^{(1,0)}(6\tau,z))^2 \\ \nonumber
& & +3\left[F^{(1,0)}(12\tau,  2z )+F^{(2,0)}(3\tau,z)+F^{(2,3)}\left( 3\tau + \frac{3}{2} ,z\right)\right],\\ \nn
{ 8A:\;}- g^{(k+2)} (\tau)\psi_1&=&  (32F^{(1,0)}(8\tau,z))^2 \\ \nn
& & +4\left[F^{(1,0)}(16\tau, 2z)
+F^{(2,0)}(4\tau,z)+F^{(2,1)}\left( 4\tau + \frac{1}{2} ,z\right)\right],\\ \nn
{ 14A:\;}-g^{(k+2)} (\tau) \psi_1&=&  (98F^{(1,0)}(14\tau,z))^2 \\ \nn
& &  +7\left[F^{(1,0)}(28\tau, 2z)+F^{(2,0)}(7\tau,z)+F^{(2,7)}\left( 7\tau + \frac{7}{2} ,z\right)\right].
\end{eqnarray}
Again using the expression for the twisted elliptic genus in (\ref{explielip}) as well as the identities (\ref{rel})  we can isolate 
the meromorphic Jacobi-form that occurs for each of these cases. 
We write this as 

\begin{eqnarray} \label{even1}
&&- g^{(k+2)}(\tau) \psi_1 = \frac{N}{2}\alpha^{(1,0)}_{g'}(\frac{N}{2}\alpha^{(1,0)}_{g'}+1)\frac{A^2(N\tau,z)}{B(N\tau,z)}+\frac{N}{2} (\alpha^{(2,0)}_{g'}-\alpha^{(1,0)}_{g'})\frac{A(\frac{N\tau}{2},z)}{B(N\tau,z)}\\ \nn
&&+N^2\alpha^{(1,0)}_{g'}A(N\tau,z)\beta^{(1,0)}_{g'} (N\tau)+\frac{N^2}{2}\beta^{(1,0)}_{g'}
(N\tau)^2 B(N\tau,z)+\frac{N}{2}\alpha^{(1,0)}_{g'} \frac{3}{16}E_4(N\tau)B(N\tau,z)\\ \nn
&&+\frac{N}{2}\left(\frac{\theta_4(N\tau,z)^2}{\theta_4^2(N\tau)}\beta^{(2,0)}_{g'}(\frac{N\tau}{2})+4\frac{\theta_2(N\tau,z)^2}{\theta_2^2(N\tau)}\beta^{(1,0)}_{g'}(2N\tau)\right)+ \phi_{g'}(\tau, z).
\end{eqnarray}
The Jacobi-form $\phi_{g'}(\tau, z) $ for each of the orbifolds is given by 
\begin{eqnarray} \label{even2}
& & { 4B}:\;\;\; \phi_{g'}(\tau, z) = \frac{ 2\theta_3^2(4\tau,z)}{\theta_3^2(4\tau)}\beta^{(2,1)}( \frac{2\tau+1}{2}),\\ \nn
& & { 6A}:\;\;\; \phi_{g'}(\tau, z) = \frac{3 \theta_3^2(6\tau,z)}{\theta_3^2(6\tau)}\beta^{(2,3)}( \frac{ 3(\tau+1)}{2} ),\\ \nn
& & { 8A}:\;\;\;   \phi_{g'}(\tau, z) =  \frac{4 \theta_3^2(8\tau,z)}{\theta_3^2(8\tau)}\beta^{(2,1)}(\frac{ 4\tau+1}{2}),\\ \nn
& & {14A}:\;\;\; \phi_{g'}(\tau, z)  = \frac{ 7\theta_3^2(14\tau,z)}{\theta_3^2(14\tau)}\beta^{(2,7)}(\frac{ 7(\tau+1)}{2} ).\\ \nn
\end{eqnarray}
From (\ref{even1})  and (\ref{even2}) we see that the only meromorphic Jacobi forms that occur in the 
expansion at $m=1$ are $A^2(N\tau, z)/ B(N\tau, z) $ and $A(N\tau/2, z)/B(N\tau, z)$. 
Therefore we can use the identity in (\ref{basicident}) to first convert the form $A(N\tau/2, z)/B(N\tau, z)$ to 
the form $A^2(N\tau, z)/ B(N\tau, z) $ and then use the the identity in (\ref{appel2}) to obtain the polar and 
the finite term.   All these manipulations can be done explicitly if necessary. 
It is clear that the mock modular form that occurs at $m=1$ in the finite part is 
again the generating function  Hurwitz-Kronecker class number ${\cal H}(N\tau, z) $.

In conclusion we have written the the Fourier-Jacobi coefficients that occur at  levels $m=0, m =1$, that is 
magnetic charge $P^2 = 0, 2$ as a polar part and a finite part for all the orbifolds in table \ref{tt}. 
We have shown that the Mock modular forms that occur in the finite term are $E_2(N\tau)$ and the 
generating function of Hurwitz-Kronecker class numbers ${\cal H}(N\tau, z)$ for these levels. 
There are identities that allow us to use the results of \cite{Dabholkar:2012nd} to obtain the finite terms 
though we are dealing with forms that meromorphic Jacobi forms that  transform under $\Gamma_0(N)$. In appendix \ref{mock}
we show that ${\cal H}(N\tau,z)$ is a Mock Jacobi form of index $1/N$ in $\Gamma_0(N)$. 
Finally we mention that just as in the $\mathbb{Z}_2$ case,  the degeneracies of 
single centered black holes can be extracted by examining Fourier coefficients of $\psi_1^F$.  
Most likely this pattern persists at higher levels in the magnetic charge expansion. It will be interesting 
to explore this further.

\section{Toroidal orbifolds} \label{sec4}

In this section we study  ${\cal N}=4$ string theories originally constructed by \cite{Sen:1995ff}. 
These involve compactification of type IIB string theory  on $T^6$ with a reflection along 
$4$ of the co-ordinates together with a $1/2$ shift along one of the remaining circles. 
The type IIA description of the theory is that of a freely acting orbifold with the action of $(-1)^{F_L}$ 
and a $1/2$ shift along one of the circles of $T^6$. For details of these two descriptions and 
the dyon configuration in this theory see \cite{David:2006ji}. 
A similar compactification but in which   the reflection in the $4$ directions  along the $T^4$ in type IIB theory is 
replaced by $2\pi/3$ rotation along one two dimensional plane of $T^4$ and a $-2\pi/3$ rotation 
along the other two dimension place together with a $1/3$ shift along one of the circles of the remaining 
$T^2$ was also discussed in \cite{David:2006ji}. 
We will call these models $\mathbb{Z}_2$ and $\mathbb{Z}_3$ toroidal orbifolds. 
The dyon partition function for these model is given by the same expression as in 
(\ref{degen}) but the Siegel form given by 
\begin{eqnarray}\label{siegform2}
\tilde{\Phi}(\rho,\sigma,v)&=&e^{2\pi i(\tilde\rho+ \tilde v)}\\ \nn
&&\prod_{b=0,1}\prod_{r=0}^{N-1}
\prod_{\begin{smallmatrix}k'\in \mathbb{Z}+
	\frac{r}{N},l\in \mathcal{Z},\\ j\in 2\mathbb{Z}+b\\ k',l\geq0, \; j<0\;  k'=l=0\end{smallmatrix}}
(1-e^{2\pi i(k'\sigma+l\rho+jv)})^{\sum_{s=0}^{N-1}e^{2\pi isl/N}c_b^{r,s}(4k'l-j^2)}.
\end{eqnarray}
Note the difference in the factor on the first line in comparison with (\ref{siegform}). 
The 
coefficients $c^{(r,s)}$ are read out from the following twisted  elliptic genus for the $\mathbb{Z}_2$ toroidal
orbifold.
\begin{eqnarray}\label{2tortwist}
F^{(0, 0)} &=& 0 , \\ \nonumber
F^{(0, 1)} &=& \frac{8}{3} A(\tau, z) - \frac{4}{3} B(\tau, z) {\cal E}_2(\tau)
, \\ \nonumber
F^{(1, 0)} &=& \frac{8}{3} A(\tau, z) + \frac{2}{3} B(\tau, z) {\cal E}_2(\frac{\tau}{2} ), 
\\ \nonumber
F^{(1, 1)} &=& \frac{8}{3}A(\tau, z) + \frac{2}{3} B(\tau, z) {\cal E}_2( \frac{\tau +1}{2} ).
\end{eqnarray}
Since this twisted elliptic genus is closely related to the $2A$ orbifold given  in (\ref{2atwist}) with $F^{(0, 0}$ set to zero
and the remaining indices multiplied by a factor of $2$. 
The corresponding Siegel 
form  is of weight $k =2$  and can be written as 
\begin{equation}\label{phitwo}
\tilde \Phi_2 (\rho, \sigma, v) = \frac{ \tilde \Phi_{6}^2(\rho, \sigma, v) }{\tilde \Phi_{10}(\rho, \sigma, v) }.
\end{equation}
Here $\tilde\Phi_6$ is the weight $6$ Siegel modular form associated with the $2A$ orbifold. 
For the $\mathbb{Z}_3$ toroidal orbifold the twisted elliptic genus is 
given by 
\begin{eqnarray} \label{3tortwist}
F^{(0, 0)} &=& 0 \\ \nonumber
F^{(0,s)}&=&A(\tau,z)-\frac{3}{4}B(\tau,z){\cal E}_3(\tau) \\ \nonumber
F^{(r,rk)}&=&A(\tau,z)+\frac{1}{4}B(\tau,z){\cal E}_3(\frac{\tau+k}{3}), \quad {r=1,2}.
\end{eqnarray}
Note  that $F^{(r, s)}$ are defined with $r, s$ mod $3$. 
The Siegel modular form associated with the the $\mathbb{Z}_3$ toroidal orbifold is of weight $k=1$ and is given by 
\begin{equation}
\tilde \Phi_1 (\rho, \sigma, v)  = \frac{\tilde\Phi_4^{3/4} (\rho, \sigma, v)}{\tilde\Phi_{10}^{1/2}(\rho, \sigma, v) }.
\end{equation}

In \cite{Sen:2010mz} it was argued that the index $d(Q, P)$ as defined in (\ref{degen}) must be positive for single 
centered black holes \footnote{ In \cite{Sen:2010mz}  $d(Q, P)$ was refered to as the index $-B_6$. }. 
The argument relied on the fact that single centered black holes are spherically symmetric and 
therefore carry zero angular momentum. The only source of signs in the index $d(Q, P)$ for 
single centered black holes then arise only from fermionic zero modes associated with a $1/4$ BPS state. 
This results in the $d(Q, P)$ being positive. 
This conjecture was verified for the orbifolds $pA$ with $p = 2, 3, 5, 7$ in \cite{Sen:2010mz}. In 
\cite{Chattopadhyaya:2017ews}  the conjecture was verified for all the orbifolds listed in table \ref{tt} which includes 
the CHL orbifolds $4B, 6A, 8A$ as well as the non-geometric orbifolds associated 
with the class $23A/B$ and classes $2B, 3B$ which lie in the Mathieu group $M_{24}$.
However it was noticed  in \cite{Chattopadhyaya:2017ews} that Siegel modular forms associated with 
 certain twisted elliptic genera written down in \cite{Paquette:2017gmb} 
did not satisfy the conjecture \footnote{See tables XVI, XVII of the Phys. Rev. D version of \cite{Chattopadhyaya:2017ews} }
These twisted elliptic genera 
satisfied the property 
\begin{equation}
\sum_{r, s = 0 }^{N-1} F^{(r, s) } = 0
\end{equation}
At present there are no known string compactification  which results in the twisted elliptic genera 
written down  by \cite{Paquette:2017gmb}  \footnote{See equations 3.20, 3.22 of Phys. Rev. D version of \cite{Chattopadhyaya:2017ews}  
for the explicit form of the twisted elliptic genera. }

In this section we show that even the toroidal $\mathbb{Z}_2, \mathbb{Z}_3$ orbifolds defined above which 
admit a   string compactification as well as a construction of dyons  do not satisfy the positivity conjecture 
for the index $d(Q, P) $ for single centered dyons at low values of charges. 
This suggests that these single centered dyons are not spherically symmetric and possibly admit hair modes 
which contain additional fermionic zero modes. 
We identify the single centered dyon by both the constraints given in \cite{Sen:2010mz} as well as subtracting the 
polar part in the meromorphic Jacobi-form that occurs at a definite magnetic charge. 

\subsection{ $\mathbb{Z}_2 $ toroidal orbifold}

The $\mathbb{Z}_2$ orbifold  compactification involves a $1/2$ shift on one of the circles of $T^2$. 
Thus the $SL(2, \mathbb{Z})$  symmetry of the torus is broken down to $\Gamma_0(2)$, which 
is now the duality symmetry of the dyon system. 
In \cite{Sen:2010mz}  the positivity property of $d(Q, P)$ was studied for for the $2A$ orbifold was studied
and the constraints on charges  for the single centered black hole was identified.
To arrive at these conditions only the $\Gamma_0(2)$ duality symmetry of the dyon system 
was used. This symmetry was used to map the wall of marginal stability that runs from 
$0$ to $i \infty$ in the axion-dilaton moduli space to all the walls which border the region ${\cal R}$
\footnote{As discussed earlier the region ${\cal R}$ in the axion-dilaton moduli space
corresponds to the contour given in equation (\ref{contour}). } (Figure \ref{cont}).
 Since the duality symmetry of the  $\mathbb{Z}_2$ toroidal orbifold  is also   $\Gamma_0(2)$, the 
conditions will remain same. 
Given the fact that the Fourier coefficients are extracted using the contour (\ref{contour}), 
the conditions on charges  that the attractor moduli lie in ${\cal R}$ 
 are given by 
\begin{eqnarray} \label{range1}
Q^2>0;\quad P^2>0; \quad Q^2 P^2>(Q.P)^2;\quad 2Q^2+P^2-3(Q\cdot P )^2\ge 0;\\ \nn
2Q^2 \ge Q\cdot P; \quad P^2 \ge Q\cdot P; \quad Q\cdot P\ge 0.
\end{eqnarray}
Later in this section  we will also  identify single centered   dyons by subtracting out the 
the polar part in the meromorphic Jacobi form that occurs at a definite magnetic charge. 
As discussed earlier, the degeneracies of single centered black holes can be obtained by 
examining the Fourier coefficients of the finite part of the meromorphic Jacobi form. 
We will also need to impost the conditions 
  in (\ref{range1})  for the Fourier coefficients so that they correspond to single centered dyons. 

We extract  the Fourier coefficients from the finite part of the meromorphic part of 
the magnetic charge expansion, or more directly using 
the contour in (\ref{contour}) with the constraints in 
 (\ref{range1}) which ensure that we are studying   single centered dyons.  We show that 
 there are cases in which 
the positivity conjecture of \cite{Sen:2010mz} is violated. 

Let us first list the some of the violations. Tables 
\ref{qp0}, \ref{qp1} \ref{qp2}  list out the degeneracy $d(Q, P) $ for small values of 
$Q^2, P^2$ with $Q\cdot P = 0, 1, 2$ respectively. 
The charges corresponding to single centered dyons in the range (\ref{range}) and violating the 
positivity conjecture are marked in bold.

\begin{table}[H] \footnotesize{
	\renewcommand{\arraystretch}{0.5}
	\begin{center}
		\vspace{0.5cm}
		\begin{tabular}{|l|c|c|c|c|c|}
			\hline
			& & & & & \\
			$Q^2\;\;\;$ \textbackslash $P^2$ & 0 & 2 & 4 & 6 & 8 \\ 
			& & & & & \\
			\hline
			& & & & &  \\
			0 & 2 & 64 & 816 & 6912 & 45584 \\
			1&-12 & {\bf -224} & {\bf -1248} & 1728 & 95104 \\
			2&48 & 1152 & 18240 & 233984 & 2432544 \\
			3&-168 & {\bf -3392} &{\bf  -10320} & 542976 & 12103360 \\
			4&528 & -11520 & 200736 & 4575744 & 86712256 \\
			5&-1512 & {\bf -30336} & {\bf -55424} & 12914944 & 412163328 \\
			6&4032 & 83968 & 1544832 & 61928448 & 2013023104 \\
			7&-10176 & {\bf -202560} & {\bf -179022} & 175358304 & 8292093664\\ 
			8&24528 & 496512 & 9480000 & 638922240 & 32998944096 \\
			9&-56796 & {\bf -1118496} & {\bf -155232} & 1735394112 & 119618619520 \\
			10&127008 & 2521600 & 49523328 & 5364983808 & 415768863360 \\
			11&-275544 & {\bf  -5374656} & 2684560 & 13858160960 & 1359548367552 \\
			12&581952 & 11389440 & 228872064 & 38347445760 & 4277873003392 \\
			13&-1199688 & {\bf -23194176} & 24502656 & 94345755264 & 12874682948352 \\
			14&2419584 & 46824960 & 959446272 & 240772494336 & 37480253184000 \\
			15&-4783968 & {\bf -91770432} & 142728318 & 566613885216 & 105389524965472 \\
			16&9288528 & 178117376 & 3712290336 & 1358448247296 & 288023853905856 \\
			17&-17735256 & {\bf  -337839744} & 678230784 & 3072125756544 & 765208401512448 \\
			18&33343344 & 634494592 & 13426540992 & 7004675317248 & 1983801614528672 \\
			19&-61794600 & {\bf -1169806144 }& 2834120592 & 15289076372544 & 5022356020513856 \\
			20&113002848 & 2136181248 & 45830851200 & 33439408301056 & 12447769083229056 \\
			21&-204081024 & {\bf -3841753664} & 10783524096 & 7071452719680 & 30229570751178240 \\
			22&364274496 & 6846494720 & 148756097664 & 149298142934016 & 72059338059045504 \\
			23&-643092768 & {\bf -12044893632} & 38123432260 & 306899147706368 & 168747648892043328 \\
			\hline
		\end{tabular}
	\end{center}
	\vspace{0.5cm}
	\caption{The index $d(Q, P) $ for  the  $\mathbb{Z}_2$ toroidal orbifold
		 some  low lying values of $Q^2$, $P^2$ with $Q\cdot P=0$ }\label{qp0}
	\renewcommand{\arraystretch}{0.5}
	}
\end{table}

\begin{table}[H] \footnotesize{
	\renewcommand{\arraystretch}{0.5}
	\begin{center}
		\vspace{0.5cm}
		\begin{tabular}{|l|c|c|c|c|c|}
			\hline
			& & & & & \\
			$Q^2\;\;\;$ \textbackslash $P^2$ & 0 & 2 & 4 & 6 & 8 \\ 
			& & & & & \\
			\hline
			& & & & &  \\
			0&	0 & 0 & {-8} & { -128} & { -1160} \\
			1&	1 & 96 & 1968 & 22528 & 190047 \\
			2&	-8 & {\bf -256} & 840 & 70912 & 1127672 \\
			3&	37 & 1376 & 34656 & 728256 & 11046139 \\
			4&	-136 & {\bf -3840} & 16632 & 2497408 & 61486056 \\
			5&	439 & 13152 & 343152 & 13144832 & 348876305 \\
			6&	-1288 & {\bf -33536} & 171152 & 42058240 & 1603241304 \\
			7&	3503 & 92928 & 2476752 & 162898624 & 7016918625 \\
			8&	-8968 & {\bf -220672} & 1265256 & 480911872 & 27503872048 \\
			9&	21854 & 540416 & 14545584 & 1556561664 & 102315259287 \\
			10&	-51080 & {\bf -1204992} & 7558560 & 4271142656 & 354800345088 \\
			11&	115154 & 2711616 & 73540080 & 12261114752 & 1175752005781 \\
			12&	-251528 & {\bf -5741824} & 38736600 & 31586749312 & 3705255587616 \\
			13&	534304 & 12144096 & 331284816 & 83106163712 & 11241057088056 \\
			14&	-1107080 & {\bf -24613888} & 176485368 & 202830655232 & 32810366529704 \\
			15&	2242936 & 49597408 & 1360242048 & 499048223424 & 92762004787995 \\
			16&	-4452488 & {\bf -96865536} & 731764656 & 1162636791680 & 254219096542800 \\
			17&	8675803 & 187681920 & 5172820416 & 2710918677760 & 678135519966520 \\
			18&	-16618760 & {\bf -355014144} & 2806978216 & 6065899132672 & 1762706150153656 \\
			19&	31335779 & 665705664 & 18435647328 & 13529566137472 & 4476930500026908 \\
			20&	-58228616 & {\bf -1224694784} & 10082072832 & 29223048194432 & 11122701903357048 \\
			21&	106740533 & 2233279616 & 62133135120 & 62776998234368 & 27083291897745248 \\
			22&	-193201800 & {\bf -4009231104} & 34221009384 & 131432145572096 & 64699862426642976 \\
			23&	345565877 & 7135993088 & 199430638848 & 273349419121472 & 151855990384385978 \\
			\hline
		\end{tabular}
	\end{center}
	\vspace{0.5cm}
	\caption{ The index $d(Q, P) $ for  the $\mathbb{Z}_2$ toroidal orbifold
		 some  low lying values of $Q^2$, $P^2$ with
	 $Q\cdot P=1$.  }\label{qp1}
	\renewcommand{\arraystretch}{0.5}
	}
\end{table}

\begin{table}[H] \footnotesize{
	\renewcommand{\arraystretch}{0.5}
	\begin{center}
		\vspace{0.5cm}
		\begin{tabular}{|l|c|c|c|c|c|}
			\hline
			& & & & & \\
			$Q^2\;\;\;$ \textbackslash $P^2$ & 0 & 2 & 4 & 6 & 8 \\ 
			& & & & & \\
			\hline
			& & & & &  \\
			0&	0 & 0 & 0 & 0 & 16 \\
			1&	0 & 0 & {- 12} & {- 224} & { -1248} \\
			2&	0 & 64 & 2592 & 43264 & 491904 \\
			3&	-2 & {\bf- 224} & 2432 & 191168 & 3805600 \\
			4&	16 & 1152 & 43392 & 1440256 & 30853488 \\
			5&	-72 & {\bf -3392} & 33720 & 5363680 & 171782688 \\
			6&	256 & 11520 & 414336 & 24533248 & 893029504 \\
			7&	-806 & {\bf -30336} & 302400 & 80281536 & 3963098880 \\
			8&	2320 & 83968 & 2926080 & 287831552 & 16432262672 \\
			9&	-6200 & {\bf -202560} & 2049968 & 851816352 & 62214237440 \\
			10&	15616 & 496512 & 16919712 & 2627695616 & 222752294016 \\
			11&	-37508 & {\bf -1118496} & 11568000 & 7176834368 & 750069187008 \\
			12&	86544 & 2521600 & 84554880 & 19942216704 & 2414262572768 \\
			13&	-192800 & {\bf -5374656} & 56838432 & 51008186976 & 7425202332576 \\
			14&	416512 & 11389440 & 377428608 & 131082715648 & 22009992439296 \\
			15&	-875808 & {\bf -23194176} & 250745920 & 317429798336 & 62951326894880 \\
			16&	1797648 & 46824960 & 1538196480 & 767174552576 & 174613994718000 \\
			17&	-3610064 & {\bf -91770432} & 1013176056 & 1773519888864 & 470403008967552 \\
			18&	7107328 & 178117376 & 5813224704 & 4077368575488 & 1234828601424128 \\
			19&	-13741542 & {\bf -337839744} & 3805021440 & 9056713382272 & 3162966840870720 \\
			20&	26130192 & 634494592 & 20608359552 & 19969539018240 & 7923569863533760 \\
			21&-48930016 & {\bf -1169806144} & 13425820256 & 42839061178880 & 19436689033887616 \\
			22&	90327040 & 2136181248 & 69137356032 & 91147913531648 & 46764751712533632 \\
			23&	-164551050 & {\bf -3841753664} & 44883305472 & 189628816546240 & 110476832098945280 \\
			\hline
		\end{tabular}
	\end{center}
	\vspace{0.5cm}
	\caption{The index $d(Q, P) $ for  the  $\mathbb{Z}_2$ toroidal orbifold
		 some  low lying values of $Q^2$, $P^2$ with $Q\cdot P=2$. }\label{qp2}
	\renewcommand{\arraystretch}{0.5}
	}
\end{table}

Let us again confirm   that the charges as well as the degeneracies  which are violating the 
positivity  conjecture for $d(Q, P)$  is that of the single centered black hole. 
For this we study the Fourier-Jacobi coefficients that occur at for low but fixed magnetic charges
but arbitrary electric charge and angular momentum. 
Thus we look for the expansion
\begin{equation}\label{exphi2}
\frac{1}{\tilde\Phi_2( q, p, y) } = \sum_{m = 0}^\infty \psi_{m}( \tau, z)  p^{\frac{m}{2} } .
\end{equation} 
Note that the expansion in terms of the magnetic variable $y$ can take 
half integral values but cannot be negative.  Here also $m$ is related to the
magnetic charge by $P^2 = 2m$. 
 From the product form of $\tilde\Phi_2$ given in (\ref{siegform})  constructed using the 
 twisted elliptic genus in (\ref{2tortwist}) we obtain
 \begin{eqnarray}\label{wtjacobitor}
[\frac{ \eta^{16}( 2\tau)} {\eta^{8}( \tau) }  ]\psi_{0} &=& -  \frac{F^{(1,0)}(2\tau,z)}{8 B(2\tau, z) }, \\ \nonumber
[\frac{ \eta^{16}( 2\tau)} {\eta^{8}( \tau) }]  \psi_{1} &=& -   \frac{F^{(1, 0)} ( 2\tau, z)^2 }{ 4 B(2\tau, z) }.
\end{eqnarray}
We define
\begin{equation}
g^{(4)} ( \tau ) = [\frac{ \eta^{16}(2 \tau)} {\eta^{8}( \tau) }].
\end{equation}
We  re-write the meromorphic Jacobi form $\psi_0$ and $\psi_1$ corresponding to 
magnetic charges $P^2 =0, P^2=2$ as a polar part and finite term. 
We perform this decomposition for $m=0$ and $m=1$. 

\vspace{.5cm}
\noindent
{\bf $P^2 = 0, m = 0$}
\vspace{.5cm}

Now we examine the case $m=0$. Using the expression for the twisted elliptic 
genus in (\ref{2tortwist}) we see that it reduces to 
\begin{equation} \label{p1tor}
- g^{(4)} (\tau) \psi_{0} =  \frac{1}{3} \frac{A(2\tau, z)}{B(2\tau, z) }
+ \frac{1}{12} {\cal E}_2( \tau) .
\end{equation}
Combining (\ref{p1}) and (\ref{p1tor}) we see that 
 the polar  and the finite part is given by 
\begin{eqnarray}
g^{(4)} (\tau)\psi_{0}^{\rm P} =  \sum_{s\in \mathbb{Z} }\frac{ q^{2s}y }{ ( 1- q^{2s} y ) ^2 }, \\ \nonumber
  \psi_{0}^{\rm F}  = \frac{1}{12 g^{(4)} (\tau) } \left(  E_2 ( 2\tau)  -  {\cal E}_2( \tau)  
\right) .
\end{eqnarray}
The finite part is certainly not  of definite sign. 
Examining the numerator we see that each of the terms are of the same sign.  This is because
\begin{equation}
 E_2 ( 2\tau)  -  {\cal E}_2( \tau) = -24  \sum_{n=1}^\infty ( \sigma( n) q^n   - \sigma( n) q^{2n}) 
\end{equation}
where $\sigma(n)$ is the divisor function of $n$. 
However due to the presence 
$g^{(4)} (\tau) $ in the denominator the signs of the terms in the $q$ expansion are not definite. 
We can contrast this situation with  the  $2A$ orbifold. The finite part for the meromorphic Jacobi form
for the $2A$ orbifold at $m=0$ is given by (\ref{2afinite}) which by the same analysis is  manifestly of a definite sign. 

As discussed for the $2A$ orbifold, 
let us also examine what is the effect of the subtraction of the polar part to the degeneracy $d(Q, P)$. 
For this recall the degeneracy is evaluated by extracting out the Fourier coefficients using the 
contour (\ref{contour}). 
This implies that we expand the polar part as follows
\begin{eqnarray}
g^{(4)} (\tau)\psi_{0}^{\rm P} &=& \frac{q^2 y}{ (1- q^2y)^2} + \frac{q^4}{ (1- q ^4 y)^2} + \cdots \\ \nonumber
&& + \frac{1}{y ( 1 - \frac{1}{y})^2} + \frac{q^2}{y ( 1- \frac{q^2}{y} )^2} + \cdots
\end{eqnarray}
This is because from the contour in (\ref{contour})  and using the 
definitions  in (\ref{fjdecomp}) we see that $ |q | <<1 $ ,  $1/|y| <<1$ and $|q^ny| <<1,  |q^n/y|  <<1$.
In each of the 
above lines we can further expand the denominators in $q$. 
Thus we see that subtraction of the polar part at $m=0$  affects only  the $d(Q, P) $ with 
$Q\cdot P \geq 1$ or  $Q\cdot P \leq - 1$ with $P^2 = 0$.  Thus the degeneracies 
with $Q\cdot P =0, P^2 = 0$ are not affected by the  subtraction of the polar part. 
This implies that the list of degeneracies in the first column of table \ref{qp0} is that 
can be obtained from the $q$ expansion of the the finite term $\Psi_{0}^{\rm F}$. This can be verified explicitly 
and indeed the terms have the sign  which is given by $(-1)^{Q^2}$. 

However note that according to the domain of charges given in (\ref{range1}) the charges which
 satisfy $Q\cdot P=0$
are not single centered. Therefore we see that the the domain (\ref{range}) is a more restrictive definition of 
single centered black holes.  
However the  Fourier coefficients of  $\psi_0^F$   is not of the same sign for the ${\mathbb Z}_2$ toroidal orbifold.

 \vspace{.5cm}
 \noindent
{\bf $P^2 = 2, m = 1$}
\vspace{.5cm}

Using the expression of the twisted elliptic genus (\ref{2tortwist}) we can write the 
  meromorphic Jacobi from that occurs at $m=1$ in (\ref{wtjacobitor}). This results in
\begin{eqnarray} \nonumber 
-  g^{(4)}  (\tau) \psi_{1} &=& 
\frac{16}{9}\frac{A^2(2\tau,z)}{B(2\tau,z)}+\frac{8}{9}A(2\tau,z){\cal E}_2(\tau)+\frac{1}{9}B(2\tau,z){\cal E}_2^2(\tau).
\end{eqnarray}
We can use the identity in (\ref{appel2}) to extract out the polar and the finite parts of $\psi_1$. 
This results in 
\begin{eqnarray}
g^{(4)} (\tau) \psi_1^{\rm P}  &=&  = 16 \sum_{n\in \mathbb{Z}}  \frac{q^{2n^2 + 2n} y^{2n+1}}{(1- q^{2n} y)^2 }, \\ \nonumber
g^{(4)}(\tau) \psi_1^{\rm F} &=& - \frac{8}{9}  A(2\tau, z) {\cal E}_2(\tau) -\frac{1}{9} B( 2\tau, z) 
\left[ E_4 (2\tau)  + {\cal E}_2^2 ( \tau)  \right]  - 32 {\cal H} ( 2\tau, z) .
\end{eqnarray}
Note again the appearance of the  function ${\cal H}$. 
Let us examine if the sign of the Fourier coefficients of the finite part $\psi_1^{\rm F} $ is the same. 
For this let again go through the analysis of what is the effect of the subtraction of the 
polar part $\psi_1^{\rm P}$.  Since we are evaluating the degeneracies in the domain (\ref{contour})
we expand the polar part as 
\begin{eqnarray}
g^{(4)} (\tau) \psi_1^{\rm P}  &=& \frac{ 16 q^{4} y^{3}}{(1 - q^2 y)^2} + 
\frac{ 16 q^{10} y^5 }{ ( 1- q^4 y)^2} + \cdots \\ \nonumber
& & + \frac{16 }{ y ( 1- \frac{1}{y} )^2} 
+ \frac{16 q^4}{ y^3 ( 1-  \frac{q^2}{y} )^2 } + \cdots
\end{eqnarray}
Therefore subtraction of the  polar part at $m=1$ affects $d(Q, P)$
with $P^2 = 2$ and $Q\cdot P \geq 3$ or $Q\cdot P \leq - 1$. 
Thus the   degeneracies in listed in the second column of 
tables \ref{qp0},  \ref{qp1} , \ref{qp2} are unchanged by the 
subtraction, 
 and therefore equal to that given by 
the finite part $\psi_1^{\rm F}$.  We have verified this explicitly, in fact this is within the domain 
given in (\ref{range}). 
Also in \cite{Dabholkar:2012nd} it was argued that the finite part captures the degeneracies of the 
single centered black hole.  
It is clear from the second column of the tables \ref{qp0}, \ref{qp1}, \ref{qp2} 
that the signs violate the positivity conjecture. 

Further more note that from the tables for $P^2 = 2$ and $Q\cdot P \leq 2$ we see that the sign
is $d(Q, P)$ is given by $(-1)^{Q^2}$. 
This can be seen also analytically by writing $\psi_1$ in terms of its product form. 
Using the form of $\psi_1$ given in (\ref{wtjacobitor}) and writing  all the modular forms in terms of their 
product representation we obtain

\begin{eqnarray}\nn
\psi_{1}=&&\frac{1}{(q_{\infty}^2)^4 qy(1-1/y)^2}\frac{\prod_{m=1}^{\infty}(1-q^{2m-1}y)^4(1-q^{2m-1}/y)^4}{\prod_{m=1}^{\infty}(1-q^{2m}y)^2(1-q^{2m}/y)^2}.\\ \label{pole}
\end{eqnarray}
Here $q^2_{\infty}=\prod_{n=1}^{\infty}(1-q^{2n})$. 
From the above equation it is evident that the sign of the coefficient of odd or even powers of $q$
is independent of the  power of $z$. This implies that no matter what is the value of $Q\cdot P$, the sign 
of the coefficient of $q^n$  is given by $(-1)^n$. 
This is what is seen in the tables \ref{qp0}, \ref{qp1}, \ref{qp2} for $ Q\cdot P = 0, 1, 2$ respectively. 
In fact this analysis shows that the positivity conjecture is violated for  infinite values of  $Q^2$ in 
with $ Q\cdot P = 0, 1, 2$  and $P^2 = 2$.  Note that these charges satisfy the condition (\ref{range}).
As argued above they are also counted 
 in $\psi_1^{\rm F}$ and therefore  correspond to single centered black holes.

The violation of the   sign of $d(Q, P)$ with   $P^2 = 2, 4$ for the case of single centered  dyons 
suggest that these states might have fermionic zero modes as hairs. This was one of the options provided 
in \cite{Sen:2010mz} if the positivity conjecture fails.  As further evidence of possible other degrees of freedom, 
we evaluate the leading saddle point statistical entropy at one loop 
 for charges satisfying the condition (\ref{range}) and therefore 
single centered.  We then compare it to the exact  degeneracy obtained by evaluating the Fourier 
coefficient $d(Q, P)$. 
Going through a similar analysis as in section \ref{sec2} for the  $\mathbb{Z}_2$ toroidal orbifold, 
we find that the statistical entropy  at one loop is given by 
\begin{eqnarray} \label{onelooptor}
S_{\rm stat}^{(1)} &=& \frac{\pi}{2\tau_2} | \frac{P}{\sqrt{2} }-  \sqrt{2} \tau  Q|^2 
- \ln g^{(4)} ( \tau) - \ln g^{(4)} (-\bar\tau)  - 4 \ln(2\tau_2) - 9 \ln 2 + O( Q^{-2}, P^{-2}) ,  \nonumber \\
\tau_1 &=& \frac{Q\cdot P}{ 2 Q^2}, \qquad \tau_2 = \frac{1}{2 Q^2} \sqrt{ Q^2 P^2 - ( Q\cdot P)^2}
\end{eqnarray}
The reason that there is a replacement of $Q \rightarrow P/\sqrt{2}$ and $P\rightarrow Q\sqrt{2}$ 
compared to (\ref{statisent1}) is that Siegel modular form $\tilde \Phi_2$ does not posses the symmetry 
in (\ref{prop1}) which is  obeyed by the modular form corresponding to the  orbifolds of $K3\times T^2$ 
Note that we have also evaluated the constant $C_1$ in the statistical entropy function. 
This was done using the relation obeyed by  $\tilde\Phi_2$  in (\ref{phitwo}). 

We now compare the one loop statistical entropy in (\ref{onelooptor}) to the exact degeneracy 
which is obtained from extracting the Fourier coefficients using (\ref{degen}) in table \ref{toro1}. 
The charges chosen are in the range (\ref{range}). Therefore they are single centered, they 
also obey the property that $d(Q, P)$ is positive. 
We note that there is a set of charges  for which the  one loop statistical entropy is 
off from the exact degeneracy by over $75 \%$. There is also a set  of charges 
for which  the statistical entropy agrees with the exact degeneracy to within  $2\%$.  
The fact that there is a set of small charges for which  the deviation from the one loop statistical 
entropy is high certainly indicates that we need to understand the geometric description for dyons in these
theories better. 

\begin{table}[H]
    \renewcommand{\arraystretch}{0.5}
    \begin{center}
        \vspace{0.5cm}
        \begin{tabular}{|c|c|c|c|c|}
            \hline
            & & & & \\
            $(Q^2,\,P^2\, Q.P)$ & $d^{\rm stat}$ & $S^{\rm stat}$ & $S^1_{\rm stat}$ & $\delta$ \\ 
            & & & &  \\
            \hline
            & & & &  \\
            (2, 4, 0) & 18240 & 9.81137 & 9.95979 & $-1.5$ \\ 
            (4, 4, 0) & 200736 & 12.2097 & 11.9331 & 2.2 \\
            (4, 8, 2) & 30853488 & 17.2448 & 17.5761 & $-1.92$ \\
            (6, 8, 1) & 1603241304 & 21.1953 & 21.5658 & $-1.75$ \\
            (8, 6, 0) & 638922240 & 20.2753 & 20.5197 & $-1.2$  \\
            \hline
  & & & &  \\
(11, 4, 0) & 2684560 & 14.80 & 18.12 & $-22.43$ \\
(3, 4, 2) & 2432 & 7.79 & 9.21 & $-18.23$ \\
(2, 4, 1) & 840 & 6.73 & 9.26 & $-37.6$ \\
(2, 8, 1) & 16632 & 9.72   & 11.6 & $19.34$ \\
(1, 6, 0) & 1728 & 7.45 & 13.26 & $-77.98$ \\
& & & & \\
\hline
        \end{tabular}
    \end{center}
    \vspace{0.5cm}
    \caption{Comparison of the statistical entropy and the statistical entropy  at one loop for the 
    toroidal ${\mathbb Z}_2$ orbifold.}
    \renewcommand{\arraystretch}{0.5}\label{toro1}
\end{table}

\subsection{ $\mathbb{Z}_3$ toroidal orbifold}

In this section we briefly discuss the $\mathbb{Z}_3$ orbifold, We first list out the violations of the positivity 
conjecture in table \ref{qp3tor}. The violations are indicated in bold face. 
Again let us mention that by performing the Fourier expansion using the 
contour in (\ref{contour}) we in the region ${\cal R}$ in the axion-dilaton moduli. 
Now demanding that the attractor moduli lie in the region ${\cal R}$ so that the dyon is single 
centered results in following constraints on the charges \cite{Sen:2010mz}. 

The condition for the dyon to be single centered are given by \cite{Sen:2010mz}. 
\begin{eqnarray}
Q^2>0, P^2 >0;  \qquad  Q^2 P^2 > (Q\cdot P)^2, Q\cdot P \geq 0,  Q\cdot P \leq  3Q^2, Q\cdot P \leq P^2, 
\\ \nonumber
5Q\cdot P \leq 6Q^2 + P^2,  \quad  5Q\cdot P \leq 3Q^2 + P^2, \quad 7Q\cdot P \leq 6Q^2 + 2P^2
\end{eqnarray}
The single centered charges that violate the positivity conjecture are given in the table \ref{qp3tor}.

\begin{table}[H]
	\renewcommand{\arraystretch}{0.5}
	\begin{center}
		\vspace{0.5cm}
		\begin{tabular}{|c|c|c|c|c|c|c|}
			\hline
			& & &  & & & \\
			$(Q^2,\;P^2)\;\;\;$ \textbackslash $Q\cdot P$ &-1 & 0 & 1 & 2 & 3 & 4 \\ 
			& & & & & & \\
			\hline
			& & & & & & \\
			(2/3, 2) &0 & 18 & 9 & 0 & 0 & 0 \\
			(4/3,2) & -252 & {\bf -36} & 45 &0 &0 &0 \\
			(4/3, 4) & -1458 & 540 & 864 & 54 & 0 & 0 \\
			(4/3, 8) & 18378 &  93816 & 72099 & 9846 & 45 & 0 \\
			(4/3,12) &-119502 &3522240 &2436363 &447606 &9243 & {\bf -6} \\
			\hline
		\end{tabular}
	\end{center}
	\vspace{0.5cm}
	\caption{Some results for the index $-B_6$ for  the torus order 3
		orbifold of $T^4$ for different values of $Q^2$, $P^2$ and $Q\cdot P$ }\label{qp3tor}
	\renewcommand{\arraystretch}{0.5}
\end{table}

For completeness we provide the meromorphic Jacobi forms that occur in the 
expansion of the inverse Siegel modular form $\tilde\Phi_1$
Again we define
\be
\frac{1}{\Phi_1(q,y,z)}=\sum_{m=0}^{\infty}\psi_m(q,z) y^{(m)/3},
\ee
where $\phi_m(q,z)$ is a modular function of 2 variables.
\begin{eqnarray}
\psi_0 &=& \frac{F^{(1,0)}}{3 B(3\tau,z)}\frac{\eta^3(\tau)}{\eta^9(3\tau)},\\ \nn
\psi_1&=&\psi_0\; (3F^{(1,0)}).
\end{eqnarray}
Rewriting these in terms in the Jacobi forms $A, B$ we obtain
\begin{eqnarray}
\psi_0=\frac{1}{3}\frac{\eta^3(\tau)}{\eta^9(3\tau)}\left(\frac{A(3\tau,z)}{B(3\tau,z)}+
\frac{1}{12}\frac{\eta^3(\tau)}{\eta^9(3\tau)}{\cal E}_3(\tau) \right), \\ \nonumber
\psi_1=\frac{\eta^3(\tau)}{\eta^9(3\tau)}\left(\frac{A^2(3\tau,z)}{B(3\tau,z)}+
\frac{1}{2}A(3\tau,z){\cal E}_3(\tau)+\frac{1}{16}B(3\tau,z){\cal E}_3^2(\tau)\right)
\end{eqnarray}
Note again since only the meromorphic forms $A(3\tau, z)/B(3\tau, z)$ and $A^2( 3\tau, z)/B(3\tau, z)$ 
occur. We can use the identities  (\ref{p2}) and (\ref{appel2}) to obtain the polar and the finite part for 
 these ratios respectively. 

\section{Conclusions} \label{sec5}

We have observed three properties of $1/4$ BPS dyons in ${\cal N}=4$ theories at low charges. 
We have seen 
that the constant $C_1$ contributes crucially to the entropy at low charges. As we have discussed
in section \ref{sec2loc}
reproducing this constant using the method of localization  proposed in 
\cite{Dabholkar:2010uh,Dabholkar:2011ec} 
will be  interesting to  pursue. In fact all the present works in this direction do not address
the $\mathbb{Z}_N$  orbifolds of $K3\times T^2$. 

We have extended the observations of \cite{Dabholkar:2012nd}. 
We showed that the for all the orbifolds considered in this paper, we can decompose the  meromorphic
Jacobi form that occurs in the 
Fourier-Jacobi decomposition of the inverse of Siegel modular forms in to a polar part and a finite part
which involves a mock modular form. This has been done to magnetic charge $P^2 = 2$. 
We have seen that the mock modular form that occurs is same as that occurred in the un-orbifolded theory 
to this order in magnetic charge. 
It will be interesting to go to higher levels in magnetic charge and see if this is always the case. 

Finally we have observed a infinite set of violations of the positivity conjecture for the Fourier coefficient
of the inverse Siegel modular form 
\cite{Sen:2010mz}. This was seen for the case of $\mathbb{Z}_2$ toroidal orbifold. We have demonstrated that the 
charges which violate the conjecture are single centered. Therefore according to the arguments of 
\cite{Sen:2010mz} it is possible that these violations might be due to the presence of hair. It will be interesting 
to understand this more precisely. Violations of the positivity conjecture was also seen 
for the $\mathbb{Z}_3$ toroidal orbifold. 

\acknowledgments

We thank Ashoke Sen for discussions which led us to evaluate the constant $C_1$. 
We thank Samir Murthy for discussions at various stages of this work and explaining 
aspects of \cite{Dabholkar:2012nd}.  We thank Boris Pioline for bringing our attention to reference
\cite{Bossard:2018rlt}, which obtains the Fourier-Jacobi coefficient for some of the CHL orbifolds at magnetic charge 
$P^2 =0$. 
We also thank  Abhishek Chowdhury, Atish Dabholkar, Rajesh Gutpa, Dileep Jatkar and Abhiram
Kidambi for discussions.  We thank the organizers of the  workshop  on `Moonshine' ,  Sep 10 to Sep 14, 2018 held at 
the Erwin Schr\"{o}dinger Institute, Vienna for a stimulating workshop which enabled us 
share preliminary results  of this work. 
The author A.C thanks CSIR for funding the research.

\appendix

\section{Details on obtaining $C_1$ from the threshold integral} \label{appen1}

In this appendix we provide the details on how to obtain the constant $C_1$ by performing the 
threshold integrals $\hat{\cal I}$ in (\ref{thresholdhat}) and $\tilde{\cal I}$ given in (\ref{threshold1}). 
The un-folding technique is used to perform these integrals. 
We can do the integral  $\tilde{\cal I}$ following \cite{David:2006ji} and we obtain the result
(\ref{tildint}). The constant arises in the integral $\hat{\cal I}$. Here 
we outline the steps in the integration which gives rise to the constant $C_1$. 
The first step to do the integral $\hat{\cal I}$ in (\ref{thresholdhat}) is to perform the
Poisson sum on $m_1, m_2$. This results in 

\begin{eqnarray}
\hat{\cal { I}}_{r,s,l} &=& \int_{{\cal F}}\frac{d^2\tau}{\tau_2^2}\frac{Y}{U_2} \sum_{r,s}\sum_{n_1, n_2, k_1, k_2 \in \mathbb{Z}} e^{-2\pi i s n_2/N} e^{2\pi i k_2r/N}\exp({\cal G}(n_1,n_2, k_1,k_2))h_l^{r,s},
\nonumber \\
\end{eqnarray}
where ${\cal G}(n_1,n_2, k_1,k_2)$ is given as,
\begin{eqnarray} \label{defcalg}
{\cal G}(n_1,n_2, k_1,k_2)=-\frac{\pi Y}{U_2^2\tau_2}|{\cal A}|^2-2\pi i \det(A)T\\ \nn
+\frac{\pi b}{U_2}(V{\tilde{\cal A}}-\bar V {\cal A})-\frac{\pi n_2}{U_2}(V^2{\tilde{\cal A}}-\bar V^2 {\cal{A}})\\ \nn
2\pi i \frac{V_2^2}{U_2^2}(n_1+n_2 \bar U){\cal A}+\frac{2\pi i\tau b^2}{4},
\end{eqnarray}
and 
\begin{eqnarray}
 A=\left( \begin{matrix}
n_1 & k_1 \\n_2 & k_2
\end{matrix}\right), \qquad  {\cal A}=\left( \begin{matrix}
1 & U
\end{matrix}\right) A \left( \begin{matrix}
\tau \\ 1
\end{matrix}\right) , \qquad   \tilde{\cal A}=\left( \begin{matrix}
1 & \bar U
\end{matrix}\right) A \left( \begin{matrix}
\tau \\ 1
\end{matrix}\right) 
\end{eqnarray}

Using the unfolding technique as in 
\cite{David:2006ji},  the integration splits into the 
 zero orbit, degenerate orbit and the non-degenerate orbit. 
 What we need to keep track of is the constants and that too the difference of the constants that 
 occur in the integral $\tilde {\cal I}$ of (\ref{threshold1}) and the integral $\hat{\cal I}$. 
 Constants can arise in the zero orbit and the degenerate orbit. 
In the zero orbit here we have $A=0$ therefore 
\[{\cal I}_{\rm zero\; orbit}=\frac{Y}{U_2}\int_{\cal F}\frac{d^2\tau}{\tau_2^2}\sum_{r,s}F^{(r,s)}(\tau,0).\]
Performing this integral, for all the orbifolds in table (\ref{tt}) we see that the constant 
that arises here is same as that of the un orbifolded $K3$ and equal to the constant
that arises in the zero orbit of the $\tilde{\cal I}$ integral. 
Lets now examine the degenerate orbit. 
Here the matrix $A$ is given by 
\begin{equation}
A=\left(\begin{matrix}
0 & k_1\\0 & k_2
\end{matrix}\right), \qquad  k_1, k_2 \in \mathbb{Z}, \quad (k_1, k_2) \neq (0, 0). 
\end{equation}
The integration region in the degenerate orbit is the strip given by 
\begin{equation} \label{strip}
-\frac{1}{2} \leq \tau_1 \leq \frac{1}{2}, \qquad  \tau_2 \geq 0
\end{equation}
Apart from the moduli dependent terms and the constant $-2k \ln{\kappa}$ as in the integral
$\tilde {\cal I}$  there is an additional constant due to the contribution 
from the twisted sector. 
The twisted sector does not contribute in the zero orbit of the $\tilde{\cal I}$ integral. 
Furthermore from  (\ref{defcalg}) the only $\tau_1$ dependence arises from the $q$ expansion of the twisted elliptic
genus. Note that since $n_2=0$, we obtain a sum over $s$. 
Then it can be seen  that the twisted elliptic genus obeys the property 
\begin{equation}
\sum_{s=0}^{N-1}F^{(r,s)}(\tau+1,z)=\sum_{s=0}^{N-1}F^{(r,s)}(\tau,z)
\end{equation}
Due to this periodicity in $\tau$, only the coefficient of $q^0$ contributes on performing the $\tau_1$ integral in 
the domain (\ref{strip}). 
Then doing the $\tau_2$ integral in the twisted sectors we are left 
  left with the  additional constant term \footnote{In principle there can be another moduli dependent term in this sector if $\sum_s c^{(r,s)}(-1)$ be non-zero. However for all orbifolds  $g'$  in table \ref{tt}  this vanishes. }
\be
C = \sum_s c^{(r,s)}(0) \sum_{k_2\in \mathbb{Z}, k_2>0}\frac{2}{k_2} e^{2\pi i k_2 r/N}.
\ee

From the explicit evaluation of the twisted elliptic genus in \cite{Chattopadhyaya:2017ews} we 
 list the values of $\sum_{s=0}^{N-1}c^{(r,s)}(0)$ for each twisted sector in different orbifolds of $K3$ 
 given in table \ref{tt}. 
 \begin{table}[H]
	\renewcommand{\arraystretch}{0.5}
	\begin{center}
		\vspace{0.5cm}
		\begin{tabular}{|c|c|c|c|c|}
			\hline
			& & & & \\
			Orbifold & Order & $k$ & $r$ & $\sum_{s=0}^{N-1}c^{(r,s)}(0)$ \\ 
			& & & & \\
			\hline
			& & & & \\
			$p$A & $p$ (prime) & $\frac{24}{p+1}-2$ & $r\ne 0$ & $k+2$ \\
			\hline
			 & & & & \\
			4B & 4 & 3 & $r=1,3$ &  4\\
			 & & & $r=2$ & 6\\
			 \hline
			 & & & & \\
			 6A & 6 & 2 & $r=1,5$ & 2 \\
			  & & & $r=2,4$ & 4 \\
			  & & & $r=3$ & 4 \\
			  \hline
			  & & & & \\
			  8A & 8 & 1 & $r=1,3,5,7$ & 2 \\
			   & & & $r=2,6$ & 3 \\
			   & & & $r=4$ & 4 \\
			   \hline
			   & & & & \\
			   14A & 14 & 0 & $r=1,3,5,9,11,13$ & 1 \\
			   & & & $r=2,4,6,8,10,12$ & 2 \\
			   & & & $r=7$ & 2\\
			   \hline
			   & & & & \\
			   15A & 15 & 0 & $r=1,2,4,7,8,11,13,14$ & 1 \\
			   & & & $r=3,6,9,12$ & 2\\
			   & & & $r=5,10$ & 2\\
			\hline
		\end{tabular}
	\end{center}
	\vspace{0.5cm}
	\caption{List of $c^{(r,s)}(0)$}\label{sumcrs0}
	\renewcommand{\arraystretch}{0.5}
\end{table}
For a prime $N$ we have,
\be
\sum_s c^{(r,s)}(0) \sum_{k_2\in \mathbb{Z}, k_2>0, r\ne 0}\frac{2}{k_2} e^{2\pi i k_2 r/N}=2(k+2)\sum_{k_2\in \mathbb{Z}, k_2>0}\frac{1}{k_2} e^{2\pi i k_2 r/N}.
\ee
We have the identity
\begin{equation}
 \sum_{k_2\in \mathbb{Z}, k_2>0}\frac{1}{k_2} e^{2\pi i k_2 r/N}=-\ln(1-e^{2\pi i r/N}).
 \end{equation}
Now summing this on the required range of $r$ ie, $0<r<n$ we get,
\be
\sum_{r=1}^{N-1}\sum_{k_2\in \mathbb{Z}, k_2>0}\frac{1}{k_2} e^{2\pi i k_2 r/N}=-\log(\prod_{r=1}^{N-1}\;(1-e^{2\pi i r/N}))=-\ln N.
\ee
Thus, we have the result
\be
C = \sum_{r=1}^{N-1}\sum_s c^{(r,s)}(0) \sum_{k_2\in \mathbb{Z}, k_2>0}\frac{2}{k_2} e^{2\pi i k_2 r/N}=-2(k+2)\ln(N).
\ee

Again from the list given in table \ref{sumcrs0} we  perform the sums in $C$ for every other orbifolds listed in 
table \ref{tt}. 
For $4B$ we have,
\begin{eqnarray}
C&=&\sum_{r=1}^{3}\sum_s 4 \sum_{k_2\in \mathbb{Z}, k_2>0, r\ne 0}\frac{2}{k_2} e^{2\pi i k_2 r/4}\\ \nn
&+&2 \sum_{k_2\in \mathbb{Z}, k_2>0}\frac{2}{k_2} e^{2\pi i k_2 /2}\\ \nn
&=& (-8\ln 4+4 \ln 2)=-10\log(4)\\ \nn
&=&-2(k+2)\ln 4.
\end{eqnarray}
In case of $6A$ we have,
\begin{eqnarray}
C&=&\sum_{r=1}^{5}\sum_s 2 \sum_{k_2\in \mathbb{Z}, k_2>0, r\ne 0}\frac{2}{k_2} e^{2\pi i k_2 r/6}\\ \nn
&+&2 \sum_{k_2\in \mathbb{Z}, k_2>0}\frac{2}{k_2} e^{2\pi i k_2 /3}+2 \sum_{k_2\in \mathbb{Z}, k_2>0}\frac{2}{k_2} e^{2\pi i k_2 /2}\\ \nn
&=& -(4\ln 6+4 \ln 3+4\ln  2)=-8\ln(6)\\ \nn
&=&-2(k+2)\ln 6.
\end{eqnarray}
Similarly for  $8A$ orbifold we get,
\begin{eqnarray}
C&=&\sum_{r=1}^{7}\sum_s 2 \sum_{k_2\in \mathbb{Z}, k_2>0, r\ne 0}\frac{2}{k_2} e^{2\pi i k_2 r/6}\\ \nn
&+&\sum_{k_2\in \mathbb{Z}, k_2>0}\frac{2}{k_2} e^{2\pi i k_2 /4}+\sum_{k_2\in \mathbb{Z}, k_2>0}\frac{2}{k_2} e^{2\pi i k_2 /2}\\ \nn
&=& -(4\ln 8+2 \ln 4+2\ln 2)=-6\ln(8)\\ \nn
&=&-2(k+2)\ln 8.
\end{eqnarray}
 Similarly one can  perform this computation for the orbifolds  $14A$ and $15A$ resulting in 
 $C= - 2(k+2) \ln 14$ and $C = -2(k+2) \ln 15$ respectively. It is interesting to note that though the 
 sums initially seems different for all the different orbifolds, they all yield $C= -2(k+2) \ln N$. 
 This concludes our derivation of the additional constant $C$ in (\ref{reshatint}). 
 
\section{Mock modular forms} \label{mock}
In this appendix we define the Mock modular forms taking the definitions from \cite{Dabholkar:2012nd} 
as follows:

\noindent
	A weakly holomorphic  pure mock modular form of weight $k\in \mathbb{Z}/2$ $h(\tau)$ is defined as:
	\begin{enumerate}
		\item $h(\tau)$ is a holomorphic function in $\mathbb{H}$ with at most exponential growth at all cusps.
		\item  The function $g(\tau)$, called the shadow of $h$, is a holomorphic modular form of weight $2-k$, and 
		\item the sum
		$\hat h := h + g^*$ is  called the completion of $h$, which transforms like a holomorphic modular
		form of weight $k$ for some congruent subgroup of $SL(2,\mathbb{Z})$.
	\end{enumerate}
	Here $g^*(\tau)$ is  the non-holomorphic Eichler integral which is the 
	 solution of the differential equation
	\be
	(4\pi\tau_2 )^k \partial_{\bar\tau}g^*(\tau)=-2\pi i\overline{g(\tau)}
	\ee

The simplest examples we encounter in the Fourier-Jacobi expansions of 
Siegel modular forms is $E_2(\tau)$ which is mock modular and  its completion $\hat E_2(\tau)=E_2(\tau)-\frac{3}{\pi\tau_2}$.
In this case we have
\be
h(\tau)=E_2(\tau),\qquad g^*(\tau)=-\frac{3}{\pi\tau_2}, \qquad \hat h(\tau)= \hat E_2(\tau).
\ee
Computing $g(\tau)$ for $E_2(\tau)$ we get $-12$ ie, a constant.
Now if we replace $\tau$ with $N\tau$ we would get,
\be
h(N\tau)=E_2(N\tau),\qquad g^*(N\tau)=-\frac{3}{\pi N\tau_2}, \qquad \hat h(N\tau)= \hat E_2(N\tau).
\ee
Now since $\hat h(\tau)$ is a modular form of $SL(2,\mathbb{Z})$ we have under a $\Gamma_0(N)$ transformation:
\be
\hat h(N\tau)\rightarrow \hat h(N \frac{a\tau+b}{cN\tau+d})\;\; {\rm with}\; ad-bcN=1,
\ee
where $a,b,c,d \in \mathbb{Z}$.
\begin{eqnarray}
\hat h(N \frac{a\tau+b}{cN\tau+d})=\hat h(\frac{a N\tau+bN}{cN\tau+d})=\hat h(\frac{a \tau'+bN}{c\tau'+d}),\\ \nn
=(c\tau'+d)^k \hat h(\tau')=(c N\tau+d)^k \hat h(N\tau).
\end{eqnarray}
with $ad-bcN=1,\; \tau'=N\tau$. Therefore  $\hat h(N\tau)$ is a modular form under  $\Gamma_0(N)$.
This observation  can be easily generalized  if $\hat h(\tau) $ itself is a modular form 
under subgroups of $SL(2,\mathbb{Z})$. 
For instance if $\hat h(\tau) $ is a modular form under $\Gamma_0(n)$, then $\hat h(\tau) $  is 
a modular form under $\Gamma_0(nN)$. 
The second example we deal with in this paper is the generating function of 
Hurwitz-Kronecker class numbers ${H}(\tau)$ which is a mock modular form 
under $\Gamma_0(4)$. Its shadow function is given by $\theta_3(2\tau)=\sum_{n\in \mathbb{Z}} q^{n^2}$. 
Therefore ${H} (N \tau)$  is a mock modular form under $\Gamma_0(4N)$ and its
shadow function is $\theta_3(2N\tau)$. 
%
%

\vspace{0.5cm}
\noindent
{\bf Jacobi Forms:} A Jacobi form $F(\tau,z)$ of weight $k$ and index $m$ is defined with the following transformation properties:
\bea \label{jac1}
F(\frac{a\tau+b}{c\tau+d},\frac{\nu}{c\tau+d}) &=& (c\tau+d)^k e^{\frac{2\pi i mc \nu}{c\tau+d}}F(\tau,\nu),\\ \label{jac2}
F(\tau, \nu+\lambda\tau+\mu) &=& e^{-2\pi i m(\lambda^2\tau+2\lambda \nu)}F(\tau,\nu).
\eea
where $\left(\begin{matrix}
a & b \\ c & d
\end{matrix}\right)\in SL(2,\mathbb{Z})$ and $(\lambda,\mu)\in \mathbb{Z}^2$

\vspace{0.5cm}
\noindent
{\bf Mock Jacobi Forms:}
By a (pure) mock Jacobi form of weight $k$ and index $m$ we denote
a holomorphic function $\phi$ on $\mathbb{H}\times C$ that satisfies the elliptic transformation property (\ref{jac2}).
A few specific properties follow for the Jacobi forms. If $F$ be a Jacobi form of weight $k$ and index $m$ then we have,
\be
F(\tau,z)=\sum_{n,r}c(4mn-r^2)q^n z^r
\ee
A weakly holomorphic Jacobi form satisfies $c(4mn-r^2)=0$, if $4mn-r^2<n_0$, if $m=1$ this is $-1$.
The Jacobi forms also satisfy:
\be
F(\tau,z)=\sum_{l\in\mathbb{Z}} q^{l^2/4m}h_l(\tau)z^l
\ee
Also this can be written as:
\be
F(\tau,z)=\sum_{l\in\mathbb{Z}/2m\mathbb{Z}}h_l(\tau)\theta_{m,l}(\tau,z),
\ee
where $\theta_{m,l}(\tau,z)=\sum_{\begin{smallmatrix}r\in \mathbb{Z}\\ r=l \mod 2m \end{smallmatrix}}q^{r^2/4m}z^r $.
For a mock Jacobi form $\phi(\tau,z)$ the modular property (\ref{jac1}) is weakened. However the completed function $\hat{\phi}$ satisfies (\ref{jac1}).
\be
\hat{\phi}(\tau,z)=\phi(\tau,z)+\sum_{l\in \mathbb{Z}/2m\mathbb{Z}} g^*_l(\tau)\theta_{m,l}(\tau,z),
\ee
with $g^*_l$ being the corresponding Eichler integral.

Since  $\hat{\phi}(\tau,z)$ satisfies (\ref{jac1}) for the full $SL(2,\mathbb{Z})$ it is easy to show that 
$\hat{\phi}(N\tau,z)$ obeys the transformation  (\ref{jac1}) for $\Gamma_0(N)$.
This is proved as follows. 
Under a modular transformation of $\Gamma_0(N)$ 
\be
\hat{\phi}(N\tau,\nu) \rightarrow \hat{\phi}(N\frac{a\tau+b}{cN\tau+d},\frac{\nu}{cN\tau+d})
\ee
where $ad-bcN=1$.
Now we know, from (\ref{jac1}) that if $a'd'-b'c'=1$, $a',b',c',d' \in \mathbb{Z}$ then,
\be
\hat{\phi}(\frac{a'\tau'+b'}{c'\tau'+d'},\frac{\nu}{c'\tau'+d'}) = (c'\tau'+d')^k e^{\frac{2\pi i m'c' \nu}{c'\tau'+d'}}\hat{\phi}(\tau',\nu).
\ee
Putting $\tau'= N\tau$ in the above equation and choosing $a'=a,b'=bN,c'=c,d'=d$ the above equation becomes:
\begin{eqnarray}
\hat{\phi}(\frac{aN\tau+bN}{cN\tau+d},\frac{\nu}{cN\tau+d})& =& (cN\tau+d)^k e^{\frac{2\pi i m'c \nu}{cN\tau+d}}\hat{\phi}(N\tau,\nu)\\ \nn
&=&(cN\tau+d)^k e^{\frac{2\pi i mNc \nu}{cN\tau+d'}}\hat{\phi}(N\tau,\nu),\quad {\rm if}\;m'=Nm.
\end{eqnarray}
Therefore we conclude that   $\hat{\phi}(N\tau,\nu)$  is a Jacobi form of index $m'/N$ and weight $k$ under $\Gamma_0(N)$ if  $\hat{\phi}(\tau,\nu)$ is a Jacobi form of index $m'$ and weight $k$.

\noindent
This implies that  completion of ${\cal H}(N\tau,z)$  denoted by  $\hat{\cal H}(N\tau,z)$ will transforms as:
\be
\hat{\cal H}(N\frac{a\tau+b}{cN\tau+d},\frac{\nu}{cN\tau+d})=(cN\tau+d)^2 e^{2\pi i \frac{c\nu}{cN\tau+d}}\hat{\cal H}(N\tau).
\ee
where, $ad-bcN=1$ and $a,b,c,d \in\mathbb{Z}$. So it is a Jacobi form under $\Gamma_0(N)$  with weight 2 and index $1/N$. 
The shadow for ${\cal H}(N\tau,z)$ will be the same as for ${\cal H}(\tau,z)$, with 
$\tau \rightarrow N\tau$.  The shadow of ${\cal H}(\tau,z)$ is given in \cite{Dabholkar:2012nd}.

{\bf Remarks:} The above analysis both in case of Mock modular and Mock Jacobi forms would go through if $b\in \mathbb{Z}/N$ and $ad-bcN=1$ and $a'd'-b'c'N=1$ and also with further restrictions on $b,b'$ being integer multiples of $M/N$ and $c,c'$ being integer multiples of some integer $M$.

\providecommand{\href}[2]{#2}\begingroup\raggedright\endgroup

\end{document}